\PassOptionsToPackage{prologue,table,xcdraw}{xcolor}

\documentclass[sigplan,screen,nonacm]{acmart}
\usepackage{graphicx}
\usepackage{xspace}

\usepackage{multirow}
\usepackage{array}
\usepackage{enumitem}
\setlist{nosep}
\usepackage{pifont}
\usepackage{subcaption}
\usepackage{algorithm}
\usepackage[noend]{algpseudocode}
\usepackage{listings}
\usepackage{cleveref}
\usepackage[font=small]{caption}
\usepackage[font=small]{subcaption}
\usepackage{pythonhighlight}
\crefname{section}{}{\S\S}
\crefdefaultlabelformat{\S#2#1#3}
\Crefformat{figure}{#2{}Figure #1{}#3}

\renewcommand\footnotetextcopyrightpermission[1]{}
\AtBeginDocument{%
  \providecommand\BibTeX{{%
    \normalfont B\kern-0.5em{\scshape i\kern-0.25em b}\kern-0.8em\TeX}}}

\setcopyright{acmlicensed}
\copyrightyear{2018}
\acmYear{2018}
\acmDOI{XXXXXXX.XXXXXXX}

\acmConference[Conference acronym 'XX]{Make sure to enter the correct
  conference title from your rights confirmation emai}{June 03--05,
  2018}{Woodstock, NY}
\acmBooktitle{Woodstock '18: ACM Symposium on Neural Gaze Detection,
 June 03--05, 2018, Woodstock, NY} 
\acmISBN{978-1-4503-XXXX-X/18/06}

\definecolor{mygreen}{rgb}{0,0.6,0}
\definecolor{mygray}{rgb}{0.5,0.5,0.5}
\definecolor{mymauve}{rgb}{0.58,0,0.82}
\newcommand{\sys}{{{Teola}}\xspace}

\newcommand{\llm}{{{LLM}}\xspace}
\newcommand{\llms}{{{LLMs}}\xspace}

\begin{document}

\title{\sys: Towards End-to-End Optimization of LLM-based Applications}

\author{
    Xin Tan$^1$, 
    Yimin Jiang$^2$, 
    Yitao Yang$^1$,
    Hong Xu$^1$
}
\affiliation{
    \institution{$^1$The Chinese University of Hong Kong, $^2$Unaffiliated}
    \city{}
    \country{}
}

\begin{abstract}
    Large language model (LLM)-based applications consist of both LLM and non-LLM components, each contributing to the end-to-end latency. Despite great efforts to optimize LLM inference, end-to-end workflow optimization has been overlooked. Existing frameworks employ coarse-grained orchestration with {task modules}, which confines optimizations to within each module and yields suboptimal scheduling decisions.
    
    We propose \textit{fine-grained end-to-end} orchestration, which utilizes \textit{task primitives} as the basic units and represents each query's workflow as a primitive-level dataflow graph. This explicitly exposes a much larger design space, enables optimizations in parallelization and pipelining across primitives of different modules, and enhances scheduling to improve application-level performance. We build \sys, a novel orchestration framework for LLM-based applications that implements this scheme. Comprehensive experiments show that \sys can achieve up to 2.09x speedup over existing systems across various popular LLM applications. The code is available at \href{https://github.com/NetX-lab/Ayo}{https://github.com/NetX-lab/Ayo}.
\end{abstract}

\maketitle
\pagestyle{plain}

\section{Introduction}
\label{sec:intro}
Large language models (LLMs) and their multi-modal variants have revolutionized user query understanding and content generation. This breakthrough has transformed many traditional and emerging applications. For instance, some search engines have integrated LLMs into their query processing pipelines, enhancing user experiences~\cite{perplexity,bing}. Additionally, AI agents, a new paradigm for human-machine interaction, have led to new applications such as emotional companionship~\cite{characterai} and personalized assistants~\cite{privatellm}.

Despite being the most intelligent component in the applications, LLMs by themselves often cannot satisfy the diverse and complicated user requirements. Examples include knowledge timeliness and long context understanding, for which LLMs cannot perform well due to their design. If not properly handled, these problems can easily cause the well-known hallucination issue~\cite{huang2023Hallucination}. To mitigate such problems, many techniques have been proposed, including RAG (Retrieval Augmented Generation)~\cite{lewis2020retrieval,liu2024iterativere}, external function calls~\cite{kim2023llmcompiler,openaifunc,huang2024tool} and even multiple LLM interactions. Popular frameworks such as Langchain~\cite{langchain} and LlamaIndex~\cite{llamaindex} support integrating various modules and building the end-to-end pipelines mentioned above.

\begin{figure}[t]
  \centering
  \includegraphics[width=0.47\textwidth]{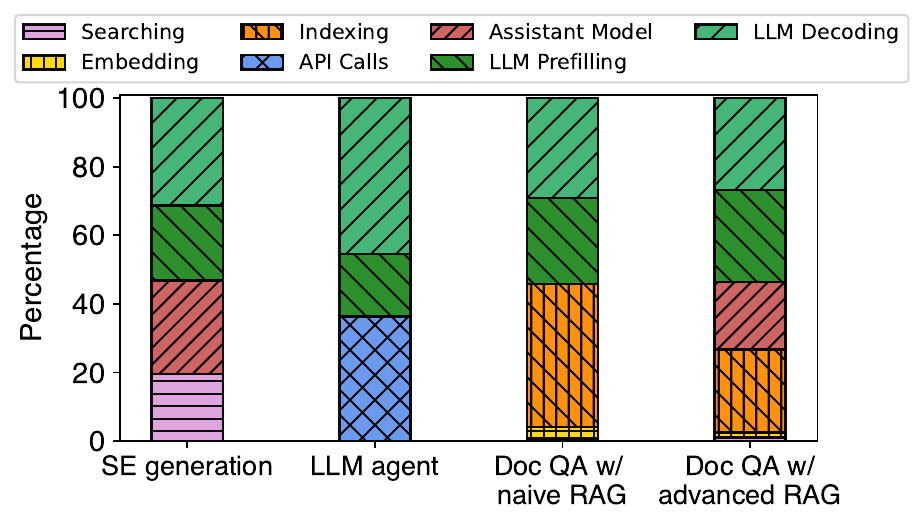} 
  \vspace{-3mm}
  \caption{Latency breakdown of each task module for various applications in Figure~\ref{fig:llm-apps} using LlamaIndex~\cite{llamaindex}. The LLM synthesizing module time is divided into prefilling and decoding.}
  \label{fig:llm_app_latency_breakdown} 
\end{figure}

\begin{figure*}[t]
  \centering
  \begin{subfigure}[t]{0.32\textwidth}
     \includegraphics[width=\linewidth]{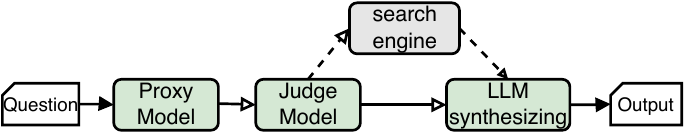}
     \caption{Search engine-empowered generation}
     \label{fig:app1}
  \end{subfigure}
  \hfill 
  \begin{subfigure}[t]{0.32\textwidth}
     \includegraphics[width=\linewidth]{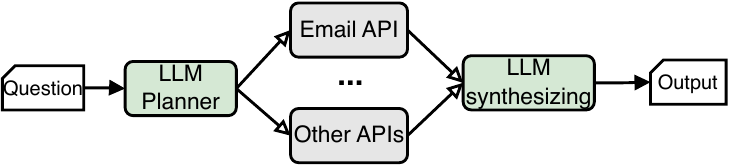}
     \caption{LLM agent with function calls}
     \label{fig:app2}
  \end{subfigure}
  \hfill
  \begin{subfigure}[t]{0.34\textwidth}
     \includegraphics[width=\linewidth]{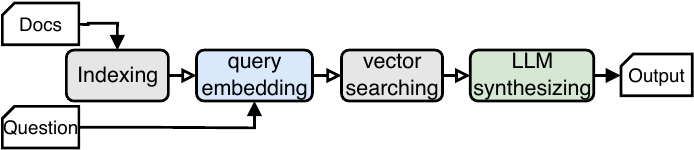}
     \caption{Document QA with naive RAG}
     \label{fig:app3}
  \end{subfigure}

  \begin{subfigure}[t]{0.485\textwidth}
     \includegraphics[width=\linewidth]{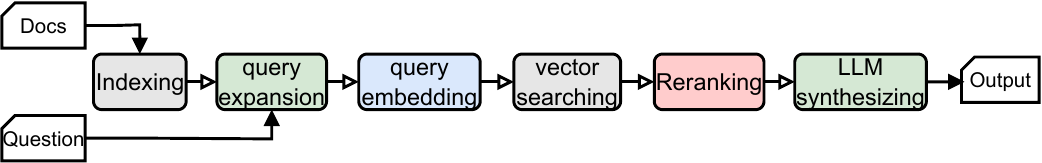}
     \caption{Document QA with advanced RAG}
     \label{fig:app4}
    \end{subfigure}
  \hfill
  \begin{subfigure}[t]{0.465\textwidth}
     \includegraphics[width=\linewidth]{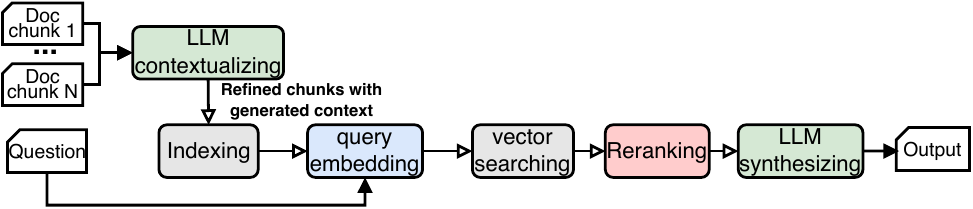}
     \caption{{Contextual retrieval from Anthropic~\cite{contextual-retrieval}}}
     \label{fig:app5}

  \end{subfigure}
  
  \caption{{Common real-world LLM-based application workflows obtained from orchestration frameworks~\cite{pairag,llamaindex,langchain} and Anthropic~\cite{contextual-retrieval}.}}
  \label{fig:llm-apps}
\end{figure*}

While significant efforts have been made to optimize LLM inference across various aspects \cite{kwon2023vllm, yu2022orca, zhong2024distserve, agrawal2023sarathi, dao2022flashattention}, little attention has been paid to the end-to-end performance of LLM-based applications composed of diverse modules. 
Figure~\ref{fig:llm_app_latency_breakdown} illustrates the execution time breakdown of several popular LLM-based applications with Llamaindex~\cite{llamaindex}. 
The non-LLM modules account for a significant portion of the end-to-end latency, and in some cases (Document question answering with RAG) even more than 50\%.
Optimizing end-to-end performance, however, faces more difficulties than one would expect in current orchestration frameworks~\cite{llamaindex,langchain,pairag,azurerag}.
They organize the workflow as a simple module-based chain pipeline (see Figure~\ref{fig:workflow_cur}), where each module independently and sequentially handles a high-level task using its own execution engines (e.g. vLLM~\cite{kwon2023vllm} for LLM inference). 
Despite its ease of use, this coarse-grained chaining scheme significantly limits the potential for workflow-level joint optimization across modules, as they treat each module as a black-box (\Cref{subsec:keyidea}).
Additionally, the decoupling of frontend orchestration and backend execution implies that request scheduling cannot optimize for the application's overall performance, forcing it to instead optimizing per-request performance, which may actually degrade the overall efficiency (\Cref{subsec:moti_application_aware}).

In this paper, we argue for a finer-grained exposition and orchestration of LLM-based applications, which can be the bedrock of end-to-end optimization. Rather than using the module-based chaining, we orchestrate with a primitive-level dataflow graph, where the \textit{task primitive} serves as the basic unit. Each primitive is a symbolic node in the graph responsible for a specific primitive operation, and has a metadata profile to store its key attributes (\Cref{subsec:keyidea}). This primitive-level graph allows us to exploit each primitive's properties and their interactions to optimize the graph, identifying an execution plan with the best end-to-end latency (\Cref{subsec:keyidea}). Furthermore, the graph captures request correlations and dependencies among different primitives as well as their topological depths, which enables application-aware scheduling and batching with better end-to-end performance (\Cref{subsec:moti_application_aware}).

Following this insight, we build \sys, a primitive-based orchestration framework for serving LLM-based application. 
\sys features two main components: 1) Graph Optimizer: It parses each user query into a specific primitive-level dataflow graph, incorporating the query's input data and configurations along with the developer's pre-defined coarse-grained workflow. Subsequently, targeted optimization passes are applied to the primitive graph to generate an efficient execution graph for runtime execution. 2) Runtime Scheduler: Utilizing a two-tier scheduling mechanism, the upper tier schedules each query's execution graph, while the lower tier is managed by individual engine schedulers. The lower tier batches and processes primitives from queries' execution graphs that request the same engine, taking into account the relationships between requests from each primitive to achieve application-aware scheduling.

We implement \sys's prototype primarily using Ray~\cite{moritz2018ray} for distributed scheduling and execution, along with various libraries for the execution engines. We evaluate \sys using diverse datasets and applications, including search engine-empowered generation, document question answering using both naive and advanced RAG, and a newly released production workflow from Anthropic~\cite{contextual-retrieval}. Comprehensive testbed experiments demonstrate that \sys can achieve up to a 2.09x speedup in end-to-end latency compared to existing schemes. These schemes include our distributed implementation of Llamaindex~\cite{llamaindex} with Ray and its advanced version that incorporates module parallelization and enhanced LLM execution, as well as another agentic framework, AutoGen~\cite{wu2023autogen}. 

Our contributions are summarized as follows:
\begin{itemize}[leftmargin=*]
\item We identify the limitations of current LLM-based orchestration frameworks, i.e. coarse-grained module-based orchestration that restricts optimization potential, and mismatch between request-level scheduling and end-to-end application performance.
\item We propose a fine-grained orchestration that represents query workflows as primitive-based dataflow graphs, enabling larger design space for end-to-end optimization including graph optimization (i.e., parallelization and pipelining) and application-aware scheduling.
\item We design and implement \sys to show the feasibility and benefit of our approach. Experiments using popular LLM applications demonstrate \sys's superior performance over current systems.
\end{itemize}

\section{Background and Motivation}

\subsection{LLM-based Applications}
\noindent\textbf{A primer on LLM.} 
Current LLMs are built upon transformers, which rely on the attention mechanism to effectively capture the long context in natural languages \cite{attention}. 
LLM inference, which this paper focuses on, is \textit{autoregressive}: in each forward pass the model produces a single new token---the basic unit of language modeling, which the becomes part of the context and is used as input for the subsequent iterations. 
To avoid redundant attention computation of preceding tokens in this process, a key-value (KV) cache is used which becomes a critical source of memory pressure \cite{yu2022orca,kwon2023vllm}.

LLM inference involves two phases: prefilling and decoding. Prefilling produces the very first output token by processing all input tokens (instruction, context, etc.), and is clearly compute-bound. 
After prefilling, the decoding phase iteratively generates the rest of the output based on the KV cache, and is memory-bound as in each iteration only the new token from the previous iteration needs to be processed. 

\noindent\textbf{LLM apps are more than just LLM.} 
Despite their great generation capabilities, \llms are not a panacea.  
Their training datasets are inevitably not up-to-date, leading to knowledge gaps and hallucination issues \cite{azurerag,lewis2020retrieval}. 
They also lack abilities to interact directly with the environment, that is they are not directly capable of sending an email though it can draft one \cite{wu2023autogen,hong2023metagpt,shen2023hugginggpt}. 
Thus real-world applications often need to integrate additional tools with LLMs to be practically usable.

{Figure~\ref{fig:llm-apps} illustrates five LLM-based application workflows, four of which are widely used in enterprise-production and open-source projects, and one recently released production workflow from Anthropic. Figure~\ref{fig:app1} demonstrates a search engine-empowered generation app, where the LLM utilizes the search engine to answer questions that are beyond its knowledge scope~\cite{bing,jeong2024adaptiverag,tan2024small}. It employs a proxy and a judge model to determine if the search engine needs to be called. 
Figure~\ref{fig:app2} illustrates a generic LLM agent, where the LLM interacts with various tool APIs to execute the formulated plan (e.g. draft and send emails using the user's account credentials)~\cite{hong2023metagpt,wu2023autogen}. 
Figures~\ref{fig:app3}~and~\ref{fig:app4} showcase document question answering (QA) with naive and advanced RAG, respectively. RAG is arguably the most popular technique to enhance LLM apps with many production uses~\cite{pairag,azurerag}. 
Here the documents uploaded by users are ingested as chunks into the vector database as the domain knowledge base~\cite{lewis2020retrieval,pairag,azurerag}, after processed by the embedding models. This step is known as indexing. The LLM uses the relevant chunks retrieved from the vector database to generate answers. The advanced version (Figure~\ref{fig:app4}) leverages LLM-based query expansion to refine and broaden new queries~\cite{jagerman2023queryexpansion,gao2023retrievalsurvey}, thereby enhancing search accuracy, and subsequently reranks all retrieved chunks for the expanded queries to synthesize a precise final answer. Lastly, Figure~\ref{fig:app5} presents Anthropic's contextual retrieval workflow~\cite{contextual-retrieval}. The LLM generates contextual summaries for each document chunk using additional content from the document. These summaries are prepended to the chunks, ensuring that each fragment maintains coherence with the original document when processed independently.

\begin{table}[]
\resizebox{0.75\columnwidth}{!}{%
\begin{tabular}{lcc}
\hline
Workflow               & \multicolumn{1}{l}{Count} & \multicolumn{1}{l}{Proportion} \\ \hline
SE generation          & 16                                   & 53.33\%                        \\
\rowcolor[HTML]{ECF4FF} 
LLM agent              & 13                                   & 43.33\%                        \\
Doc QA w/ naive RAG    & 26                                   & 86.67\%                        \\
\rowcolor[HTML]{ECF4FF} 
Doc QA w/ advanced RAG & 23                                   & 76.67\%                        \\ \hline
\end{tabular}%
}
\vspace{2mm}
\caption{{The occurrence count and proportion of the four workflows (Figure~\ref{fig:llm-apps}(a)--(d)) in 30 sampled GitHub projects with more than 1K stars.}}
\label{fig:llm_app_prevelance} 
\vspace{-5mm}
\end{table}

\begin{figure}
    \centering
    \includegraphics[width=\columnwidth]{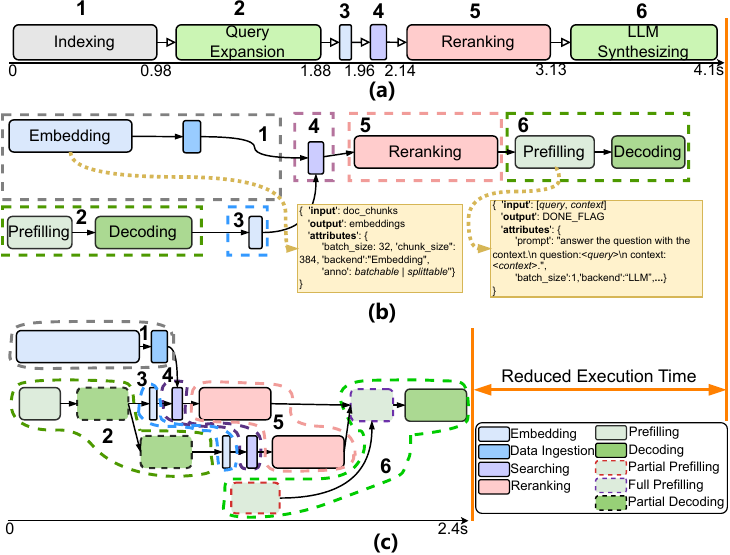}
    \vskip -0.2in
\captionsetup[subfigure]{labelformat=empty}
 \begin{minipage}[]{0\linewidth}
 \subcaption{}
\label{fig:workflow_cur}
\centering
\end{minipage}\hfill
\begin{minipage}[]{0\linewidth}
\centering
\subcaption{}
\label{fig:workflow_our} 
\end{minipage}\hfill
\begin{minipage}[]{0\linewidth}
\centering
\subcaption{}
\label{fig:workflow_our_exec}
\end{minipage}
\caption{Workflow expression and execution comparison of existing schemes and \sys. (a) Module-level workflow in current schemes. Note the arrows here merely indicate the execution order, not the dependency. %
(b) Primitive-based dataflow graph in \sys (limited metadata of nodes shown in the interest of space). Arrows here represent the dependency between primitives.
(c) Execution graph after optimization in \sys.
}
\label{fig:workflow_com}
\end{figure}

We also conduct a small case study to validate the prevalence of the first four workflows by searching GitHub for open-source projects using the keyword ``LLM applications.'' 
We select 30 best-matched projects with more than 1,000 stars, including notable ones such as Haystack~\cite{haystack}, AutoGen~\cite{wu2023autogen}, and Langchain~\cite{langchain}. 
We manually check these projects to see if the four workflows or their similarities were present; it is common for them to contain multiple workflows. The results, presented in Table~\ref{fig:llm_app_prevelance}, illustrate the popularity of these LLM-based workflows. 
We observe that most workflows occur in more than half of the sampled projects, with RAG-based QA being the most popular in current open-source projects.

With the increasing adoption of such workflows, it is evident that the LLM may not be the sole performance bottleneck in complex application pipelines, as discussed earlier in \Cref{sec:intro}. Therefore, carefully orchestrating the various components of the workflow is critical.
}

\subsection{Fine-grained Orchestration of LLM Apps}
\label{subsec:keyidea}

Several frameworks, such as LlamaIndex~\cite{llamaindex}, Langchain~\cite{langchain}, and enterprise solutions like PAI-RAG~\cite{pairag} and Azure-RAG~\cite{azurerag}, have emerged to facilitate the creation and orchestration of LLM applications.
They naturally adopt \textit{module-level} orchestration, where each app is defined and executed as a simple chain of modules, as depicted in Figure~\ref{fig:workflow_cur}. Each module is executed independently on their backend engines. 
While coarse-grained module-level chaining is easy to implement and use, it is inherently limited in optimizing complex workflows for performance.
It overlooks the larger design space of jointly optimizing the modules, particularly by exploiting the intricate dependencies among the internal operations of the individual modules. 
Moreover, several agentic-like frameworks~\cite{wu2023autogen,autogpt} feature a more compact and coarse-grained pattern, in which multiple agents with distinct roles working on distinct sub-tasks are connected with a predefined workflow to complete a complex task. 
These frameworks have not explored opportunities of end-to-end optimization across agents or sub-tasks either.

The central thesis of this paper is to advocate for fine-grained exposition and orchestration of \llm apps in order to improve end-to-end performance. 
Consider an alternative representation of the same app workflow (Figure~\ref{fig:workflow_cur}) shown in Figure~\ref{fig:workflow_our}.
Instead of working with modules, we decompose each module into fine-grained \textit{primitives} as the basic unit of orchestration (i.e. nodes in the graph). 
The indexing module, for example, is decomposed into embedding creation and data ingestion primitives, and query expansion is decomposed into prefilling and decoding primitives just like the LLM synthesizing module.

Moreover, the dependency among these primitives are explicitly captured in this dataflow graph, enabling the exploration of more sophisticated joint optimizations across primitives and modules. 
As a simple example, it is apparent that the embedding creation and data ingestion primitives can be executed in parallel with prefilling and decoding primitives for query expansion in Figure~\ref{fig:workflow_our}, since their inputs are independent.
Each primitive's input/output relationship, along with other key information (see \Cref{subsec:moti_application_aware} for some examples), is encoded as node attributes in the graph. 

We can then optimize this primitive-level dataflow graph to identify the best execution plan of the entire workflow with the lowest end-to-end latency.
Figure~\ref{fig:workflow_our_exec} shows the optimized execution graph corresponding to the dataflow graph in Figure~\ref{fig:workflow_our}. Specifically, given that query expansion creates multiple new queries, the corresponding decoding primitive can run in a pipeline fashion with multiple \textit{partial decoding} primitives, each generating a new query and sending it to the subsequent primitive (embedding creation) right away without waiting for all queries to come out.
By the same token, the prefilling primitive of LLM synthesizing can be divided into a partial prefilling first that operates on the system instruction and user query, which can run in parallel with indexing before search and reranking.

Thus, primitive-level dataflow graph allows us to explore various parallelization and pipelining opportunities across primitives that are not visible in existing module-based orchestration.
The gain is significant: in our example (Figure~\ref{fig:workflow_com}), the overall execution time is reduced from 4.1s to 2.4s.

\subsection{Application-Aware Scheduling and Execution}
\label{subsec:moti_application_aware}

\begin{figure}[t]
  \centering
  \includegraphics[width=0.495\textwidth]{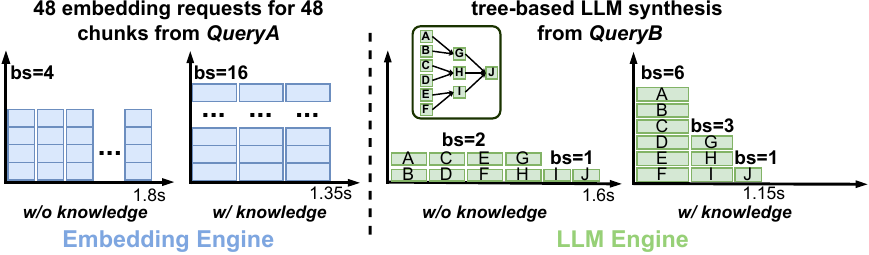} 
   \begin{minipage}[t]{0.545\linewidth}
        \vspace*{-13pt}
        \subcaption{Batching for embedding engine.}
        \label{fig:embedding_batch}
      \end{minipage}
      \hfill
      \begin{minipage}[t]{0.43\linewidth}
      \centering
        \vspace*{-13pt}
        \subcaption{Batching for LLM engine.}
        \label{fig:llm_batch}
      \end{minipage}
 \caption{Comparison between request-level and application-level scheduling and execution.}
\end{figure}

Another limitation of current \llm application orchestration is the request-level optimization of the backend execution engines, which is a mismatch with the application-level performance that the user perceives.  
Suppose that Triton~\cite{tritonserver}, a popular serving engine, is used to serve the embedding model for the indexing module. The Triton engine treats each request uniformly with a fixed batch size of 4 as shown in Figure~\ref{fig:embedding_batch}.
Without any application-level information, we can only optimize for the per-request latency with reasonable but sub-optimal GPU utilization.

Now given that these embedding requests come from the same module, it is obvious that the execution engine should optimize for the total completion time instead of per-batch latency.
Therefore, a better strategy is to use a larger batch size of say 16 to fully utilize the GPU.
With 48 requests in total (for 48 document chunks), the total completion time is reduced from 1.8s to 1.35s, a 1.3x speedup as seen in Figure~\ref{fig:embedding_batch}, even though the per-batch latency is slightly higher.

The above toy example utilizes request correlation of an individual primitive. 
Another type of information we can exploit is request dependency across primitives. %
Consider the LLM synthesizing module, which makes a series of LLM calls in a tree-based synthesis mode in Figure~\ref{fig:llm_batch}. 
These requests are executed with a batch size of 2 in conventional request-level scheduling, again to optimize per-request latency. 
In contrast, given that they form a dependency tree of depth of 2, the LLM execution engine can process requests at the same depth with varying batch sizes, leading to a 1.4x speedup overall albeit longer per-batch latency.

To sum up, fine-grained orchestration also bridges the gap from the execution engine's request-level optimization and enables application-aware scheduling, by using request correlation and dependency information from the primitive dataflow graph (as node attributes) to further optimize end-to-end performance.

\section{Design Overview}
\label{sec:design}
\begin{figure}[t]
  \centering
  \includegraphics[width=0.485\textwidth]{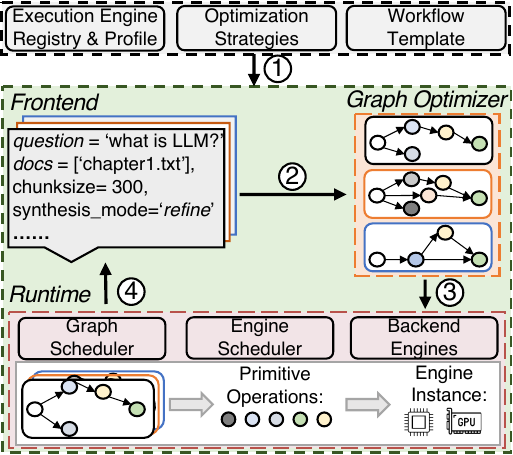} 
  \caption{System Overview of \sys.}
  \label{fig:sys_overview} 
\end{figure}

\subsection{Architecture}
\label{subsec:arch}

\sys is a novel orchestration framework to optimize the execution of LLM-based applications with primitive operations as the basic unit.

Figure~\ref{fig:sys_overview} depicts \sys's architecture. 
In the offline stage \ding{172}, developers register execution engines for an app, such as those for embedding models, LLMs, and database operations, along with their latency profiles for various input sizes (e.g., batch size and sequence length).
They also provide a workflow template that outlines the app's components (e.g., query expansion and LLM generation) and their execution sequence, similar to tasks modules in current frameworks~\cite{llamaindex,langchain}.
Optionally, developers may specify optimization strategies for certain primitive operations.
Once the app is configured and deployed, the system is ready for online serving.

In the online stage, upon receiving a query with specific input data and workflow configurations, \sys creates a primitive-based dataflow graph (\textit{p-graph}) \ding{173}, applies relevant optimizations to generate the execution graph (\textit{e-graph}), and submits the e-graph to the runtime \ding{174}. The runtime accurately tracks and efficiently schedules the execution of the e-graph's primitives on the appropriate backends. Finally, the results are returned to the frontend upon completion \ding{175}.

\subsection{APIs}
\label{subsec:api}

Listing~\ref{lst:code} presents a simplified usage example of \sys, highlighting its main components as described below.

\noindent\textbf{Execution engines.} Execution engines handle requests for models or operations from workflow components (line 5). They can be model-free or model-based. Model-free engines, such as databases, are primarily CPU-based and do not involve DNN models. On the other hand, model-based engines can deploy various DNN models, including BERT-family models~\cite{devlin2018bert} for embedding and LLMs for generation. A single engine can serve multiple components with different purposes, such as the shared LLM engine for query expansion and LLM synthesizing in Figure~\ref{fig:app4}.

\begin{lstlisting}[float=t,language=python,caption={Simplified usage example of \sys.},basicstyle=\scriptsize\ttfamily,,label={lst:code}, breaklines=true,captionpos=b]
from Teola.app import APP,Node
from Teola.executor import Engine
from Teola.graph import OPT_Pass,Graph
# Register executor engines (e.g. LLM).
LLM=Engine("LLM",executable=llm_exec,config=config_llm,resource={"GPU":2},instances=2)
# Register optimization passes.
OPT_Pass.register("pipelining",pipline_pass)
app=APP.init() # Init an application 
# Resiger template components with specification
query_expand=Node("LLM",in_kwargs,out_kwargs, 
    anno="splitable",config=config_expand)
embedding=Node("Embedding",in_kwargs,out_kwargs, 
    anno="batchable",config=config_embed)
# Omit other components...
generation=Node("LLM",in_kwargs,out_kwargs,
    anno=None,config=config_gen)
# Declare the dependency.
query_expand>>embedding>>...>>generation
# Update the app's workflow template.
app.update_template([query_expand,...])
# Construct and optimize graph based on query and config.
e_graph=Graph.optimize(app,query,config,OPT_Pass)
# Sumbit to runtime and schedule.
e_graph.schedule()

\end{lstlisting}

\noindent\textbf{Workflow definition.} While \sys abstracts away the complexity of graph optimization and scheduling, developers must first define a high-level workflow template with runtime-rendered placeholders as input for each application, similar to existing frameworks~\cite{langchain,llamaindex,promptflow,wu2023autogen}. 
Using \sys's APIs (lines 8-20), developers specify essential components with their required engines, roles, and input-output configurations. These components can be annotated with optimization hints, such as \texttt{batchable} (for independent inputs) and \texttt{splittable} (for divisible outputs). The $>>$ operator establishes execution sequences across components to ensure dataflow correctness. 
This workflow template serves as the foundation for constructing and optimizing detailed execution graphs that adapt to various query configurations.

\noindent\textbf{Graph optimization.} For each query, a finer-grained p-graph with primitive nodes is constructed based on the query-specific data, configuration, and predefined workflow template (line 22). \sys then utilizes built-in optimization passes for primitive operations and patterns to identify an optimized execution plan and generate an e-graph (\Cref{sec:graph_opt}) for execution. Developers can also register custom optimizations through a provided interface (line 7).

\noindent\textbf{Declarative query.} A declarative interface is offered for submitting queries to a deployed app. Beyond specifying queries (i.e., question and context), users can customize the workflow (Figure~\ref{fig:sys_overview}), allowing for parameter tuning of components (e.g., document chunk size for indexing, LLM prompt template and LLM synthesis mode) to meet performance expectations.

\section{Graph Optimizer}\label{sec:graph_opt}

Graph optimizer generates a fine-grained, per-query representation (\textit{p-graph}) by combining query information and workflow template. This p-graph, composed of symbolic primitive nodes, enables optimization strategies to produce an efficient \textit{e-graph} for execution.

\subsection{p-Graph}

\noindent\textbf{Primitives.} Relying solely on high-level components, as discussed in \Cref{subsec:keyidea}, can oversimplify the intricate relationships between operations and expose limited information and flexibility. To address this, we introduce a refined abstraction: the task primitive (primitive for short). Akin to the operation nodes in TensorFlow~\cite{tensorflow}, symbolic primitives at the workflow level enhance granularity in representation and provide valuable information for optimization prior to execution.

Specifically, as shown in Table~\ref{tab:TPr_example}, a primitive can correspond to the functionality of a standard operation within a registered execution engine (e.g., embedding creation in embedding engines or context ranking in reranking engines) or represent a fine-grained decomposed operation. 
For instance, LLM inference is decomposed into \texttt{Prefilling} and \texttt{Decoding}, with \texttt{Partial Prefilling} and \texttt{Full Prefilling} constituting LLM prefilling, and \texttt{Partial Decoding} managing different parts of full decoding. Additionally, primitives can be control flow operations such as aggregation or conditional branching (i.e., \texttt{Aggregate} and \texttt{Condition}). 
Each primitive includes a metadata profile detailing its inputs, outputs, parent nodes, and child nodes, forming the basis for graph construction. This profile also contains key attributes such as batch size for DNNs or prompts for LLMs, as well as the target execution engine.

\begin{table}[]
\resizebox{\columnwidth}{!}{%
\begin{tabular}{ll}
\hline
Type & Description \\ \hline
Reranking & \begin{tabular}[c]{@{}l@{}}Compute and rank the relevance scores for the query \\ and context pairs.\end{tabular} \\
Ingestion & Store embedding vectors into vector database \\
Searching & Perform vector searching in the database \\
Embedding & Create embedding vectors for docs or questions \\
\rowcolor[HTML]{ECF4FF} 
Prefilling & The prefilling part of LLM inference \\
\rowcolor[HTML]{ECF4FF} 
Decoding & The decoding part of LLM inference \\
\rowcolor[HTML]{ECF4FF} 
Partial Prefilling & \begin{tabular}[c]{@{}l@{}}Prefilling for partial prefix of a prompt (e.g. instruction, \\ context, question)\end{tabular} \\
\rowcolor[HTML]{ECF4FF} 
Full Prefilling & Prefilling for rest part of a prompt after a partial prefilling \\
\rowcolor[HTML]{ECF4FF} 
Partial Decoding & Part of full decoding for partial output \\
\rowcolor[HTML]{EFEFEF} 
Condition & Decide the conditional branch \\
\rowcolor[HTML]{EFEFEF} 
Aggregate & Aggregate the results from multiple primitives \\ \hline
\end{tabular}%
}
\caption{Primitive examples in Figure~\ref{fig:app4}. White backgrounds denote common operations, blue for decomposed operations, and gray for control flow operations.}
\label{tab:TPr_example}
\end{table}

\noindent\textbf{p-Graph construction.} The optimizer converts the original workflow template $\mathcal{T} = (\mathcal{T}_N, \mathcal{T}_E)$ with query-specific configuration $\mathcal{C} = (\mathcal{T}_N, \mathcal{C}_N)$ into a more granular p-graph $\mathcal{G} = (\mathcal{V}_N, \mathcal{V}_E)$ as outlined in Algorithm~\ref{algo::dag_trans_opt}, where $\mathcal{T}_N$ represents components, $\mathcal{T}_E$ dependencies, and $\mathcal{C}_N$ user configurations. The process decomposes each template component into explicit symbolic primitives based on the configuration, creating a sub-primitive-level graph with well-defined dependencies. For instance, the LLM synthesizing module in \textit{refine} mode with 3 context chunks is transformed into a sub-graph where 3 pairs of \texttt{Prefilling} and \texttt{Decoding} primitives are chained and configured with corresponding metadata. The final resulting p-graph preserves the original workflow dependencies while providing a more detailed view of the workflow's inner workings.

\begin{figure*}[t]
  \centering
  \includegraphics[width=\textwidth]{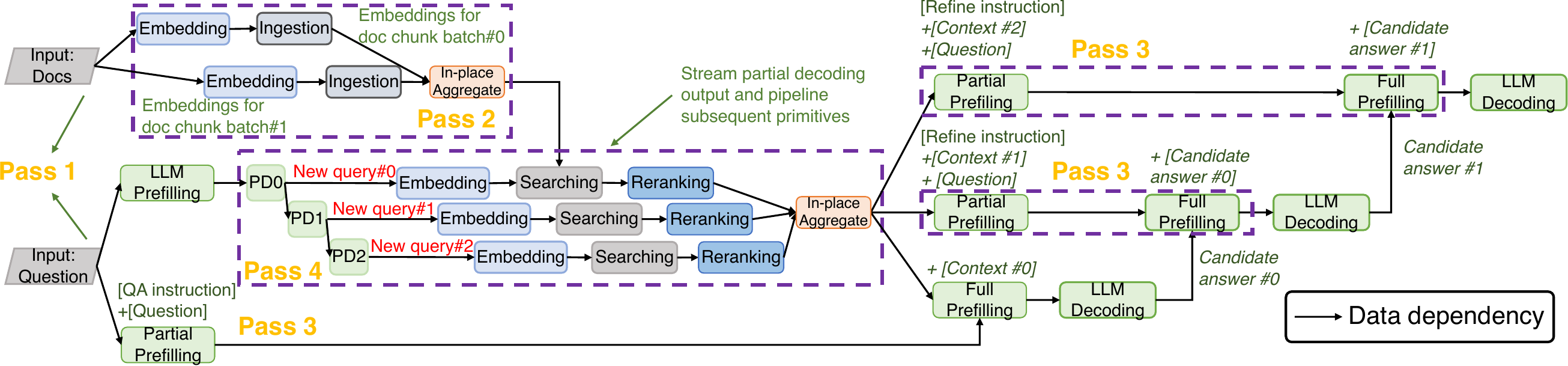} 
  \caption{An illustrative optimized e-graph of a query for the advanced RAG-based document QA with a \textit{refine} synthesis mode. (PD: partial decoding; annotated computed prompt part for {Partial/Full Prefilling}; primitive metadata omitted; block length not indicative of execution time.)}
  \label{fig:opt_graph} 
\end{figure*}

\begin{algorithm}[t]
\footnotesize
\caption{Graph transformation and optimization}\label{algo::dag_trans_opt}
\begin{algorithmic}[1]

\Statex 
\Function{GraphTransform}{$\mathcal{T}, \mathcal{C}$}
    \State $\mathcal{V}_N \gets \{\}$; $\mathcal{V}_E \gets \{\}$ \Comment{primitives and their data dependency into}
    
   \Statex\quad\quad{\textcolor{blue}{\# \textit{Decompose each template component with configuration into}}}
   \Statex\quad\quad{\textcolor{blue}{\# \textit{a sub-graph with primitives and maintain sub-graph dependency}}}
    \For{each $t \in \mathcal{T}_N$}
        \State $Prims,Edges \gets \texttt{DecomposeComponent}(t, \mathcal{C})$ 
        \State $Prims \gets \texttt{Configure}(Prims,\mathcal{C})$
        \State $\mathcal{V}_N.\text{extend}(Prims)$; $\mathcal{V}_E.\text{extend}(Edges)$
        
    \EndFor
\Statex\quad\quad{\textcolor{blue}{\# \textit{Maintain template's original component dependency }}}
    \For{each $(t_i, t_j) \in \mathcal{T}_E$}
        \State $tailp \gets \texttt{GetTailPrim}(t_i)$; $headp \gets \texttt{GetHeadPrim}(t_j)$
        \State $\mathcal{V}_E.\text{append}((tailp, headp))$
    \EndFor

    \State \Return $\mathcal{G}_p = (\mathcal{V}_N, \mathcal{V}_E)$\Comment{return primitive-level \textit{p-graph}}
\EndFunction

\Function{GraphOpt}
{$\mathcal{G}_{p}$,$\mathcal{P}$}
 \Statex\quad\quad{\textit{\textcolor{blue}{\# $\mathcal{G}_{p}$ for p-graph, $\mathcal{P}$ for profile of execution engines}}}
   
    \State $\mathcal{G}_{e} \gets \texttt{PrunDependency}(\mathcal{G}_{p})$ \Comment{Pass 1}

    \State $\mathcal{G}_{e} \gets \texttt{StageDecompose}(\mathcal{G}_{e},\mathcal{P})$ \Comment{Pass 2}

    \State $\mathcal{G}_{e} \gets \texttt{PrefillingSplit}(\mathcal{G}_{e})$ \Comment{Pass 3}

     \State $\mathcal{G}_{e} \gets \texttt{DecodingPipeling}(\mathcal{G}_{e})$ \Comment{Pass 4}

    \State \Return $\mathcal{G}_e$\Comment{return optimized \textit{e-graph}}

\EndFunction
\end{algorithmic}
\end{algorithm}

\subsection{Optimization}\label{subsec:opt_example}

As mentioned in \Cref{subsec:keyidea}, \sys focuses on maximizing parallelism in distributed execution rather than single-point optimization or acceleration (orthogonal and discussed in \Cref{sec:related_work}). Specifically, the optimizer identifies opportunities for primitive parallelism (parallelization) and pipeline parallelism (pipelining), employing a set of static, rule-based optimizations.

\noindent\textbf{Exploitable opportunities.} Firstly, the original dependencies inherited from the workflow template, which only depict a high-level sequence of components, may introduce redundancy in the fine-grained p-graph. 
To maximize parallelization, it is essential to analyze and prune unnecessary dependencies, thereby freeing independent primitives and creating parallel dataflow branches (\textbf{Pass 1}). Additionally, compute-intensive primitives can be broken down into multiple pipelining stages, where feasible, enabling them to be executed concurrently with subsequent primitives (\textbf{Pass 2}).

Furthermore, we have observed that the core of the workflow, the LLM, has exploitable special attributes. Specifically, two key attributes can be leveraged: (1) \textit{causal prefilling}: This allows the LLM's prefilling to be split into dependent parts, enabling parallelization of partial prefilling with preceding primitives (\textbf{Pass 3}), and (2) \textit{streaming decoding output}: The auto-regressive and partial output of specific LLM decoding can be pre-communicated as input to the downstream primitives, creating additional pipelining opportunities (\textbf{Pass 4}).

\noindent\textbf{Optimization passes.}
Based on the above analysis, the following optimization passes are integrated and can be applied to the p-graph to optimize end-to-end workflow execution:
\begin{enumerate}[leftmargin=*,label=\(\triangleright\)]
\item\noindent{\textit{Pass 1: Dependency pruning.}} To increase parallelization potential, we eliminate unnecessary dependencies and identify independent dataflow branches for concurrent execution by examining each task primitive's inputs with its current upstream primitives. Redundant edges are pruned, ensuring that remaining edges represent only data dependencies, which may detach certain task primitives from the original dependency structure. For example, primitives in query expansion and embedding modules are detached to form a new branch in Figure~\ref{fig:workflow_our_exec}.

\item\noindent\textit{Pass 2: Stage decomposition.} For \texttt{batchable} primitives that process data exceeding the engine's maximum efficient batch size (i.e., the size beyond which throughput does not increase), they are decomposed into multiple stages, each handling a sub-micro-batch and pipelining with downstream \texttt{batchable} primitives. While more aggressive division may increase pipelining degree, finding the optimal split size is time-consuming. Moreover, adjacent batchable primitives lead to an exponential search space, which is impractical for latency-sensitive scenario. To balance resource utilization and execution efficiency, we only explicitly segment a primitive into multiple stages when its input size reaches the maximum efficient batch size. An \texttt{Aggregate} primitive is added at the end of pipelines to explicitly synchronize and aggregate the results if necessary.

\item \noindent\textit{Pass 3: LLM prefilling split.} In LLM prefilling, a full prompt consists of components such as system/user instructions, questions, and context. Within a workflow, some prompt parts may be available in advance (e.g., user instructions or questions), while others may not be (e.g., retrieved context from database in RAG). Instead of waiting for all components, available components of a prompt can be opportunistically pre-computed as they become ready, while respecting the causal attribute of attention computation, thus enabling partial prefilling parallelization.

\item \noindent\textit{Pass 4: LLM decoding pipeling.} During the decoding process of LLM, tokens are generated incrementally. Once a coherent output (e.g., a new rewritten sentence in query expansion) is available, it can be promptly forwarded to downstream \texttt{batchable} primitives, avoiding delays associated with waiting for full decoding. To enable this optimization, the LLM call must be annotated as \texttt{splittable}, indicating that its outputs can be semantically divided into distinct parts. The corresponding parser monitors the progressive, structured output (e.g., JSON) of the decoding process, extracting and forwarding complete pieces of a partial decoding to successors as soon as they become available.

\end{enumerate}

\noindent\textbf{Optimization procedure.} The optimizer iteratively traverses the p-graph, matching primitive nodes to the pattern of each optimization pass. When a match is found, the corresponding pass is applied, and the relevant primitives are modified accordingly, as outlined in Algorithm~\ref{algo::dag_trans_opt}. This process continues until no further optimizations are possible. To reduce overhead, a cache can be employed to store and reuse the results of optimized subgraphs.

\noindent\textbf{An example.} Figure~\ref{fig:opt_graph} presents an optimized e-graph for the app example shown in Figure~\ref{fig:app4}. In this optimized e-graph, query expansion module generates three new queries to enhance searching process, and the top three retrieved chunks are fed into the LLM synthesizing module. The LLM synthesizing module operates in \textit{refine} mode, first generating an initial answer using the top chunk with a QA-style prompt template. It then refines the candidate answer twice using the remaining two chunks with a \textit{refine}-style prompt template. The different passes applied by the optimizer are annotated in the figure for clarity.

\section{Runtime Scheduling}\label{sec:runtime}

\sys utilizes a two-tier scheduling mechanism at runtime. The upper-tier graph scheduler dispatches primitive nodes of each query's optimized e-graph. The lower tier consists of engine schedulers that manage engine instances and fuse primitive requests from queries for efficient execution. Separating graph scheduling and operation execution enhances scalability and extensibility for \sys.

\subsection{Graph Scheduler}

The graph scheduler closely tracks the status of each query's e-graph and issues primitive nodes as their dependencies are met. It evaluates node in-degrees and dispatches nodes to the appropriate engine scheduler when in-degrees reach zero. Note that the graph scheduler dispatches the node itself rather than its associated requests, ensuring that the lower scheduler can identify requests originating from a primitive, instead of treating them independently like in existing frameworks (see ~\Cref{subsec:moti_application_aware}). Upon completion of a primitive's execution, the scheduling thread is notified via RPC calls, and the output is transferred. The thread then decrements the in-degrees of downstream primitives, preparing them for execution.

Additionally, a dedicated per-query object store manages intermediate outputs. This store acts as both an input repository for pending primitives and offers a degree of fault tolerance, safeguarding against operation failures.

\begin{figure}[t]
    \centering
      \begin{subfigure}{\columnwidth}
        \centering
        \includegraphics[width=\columnwidth]
    {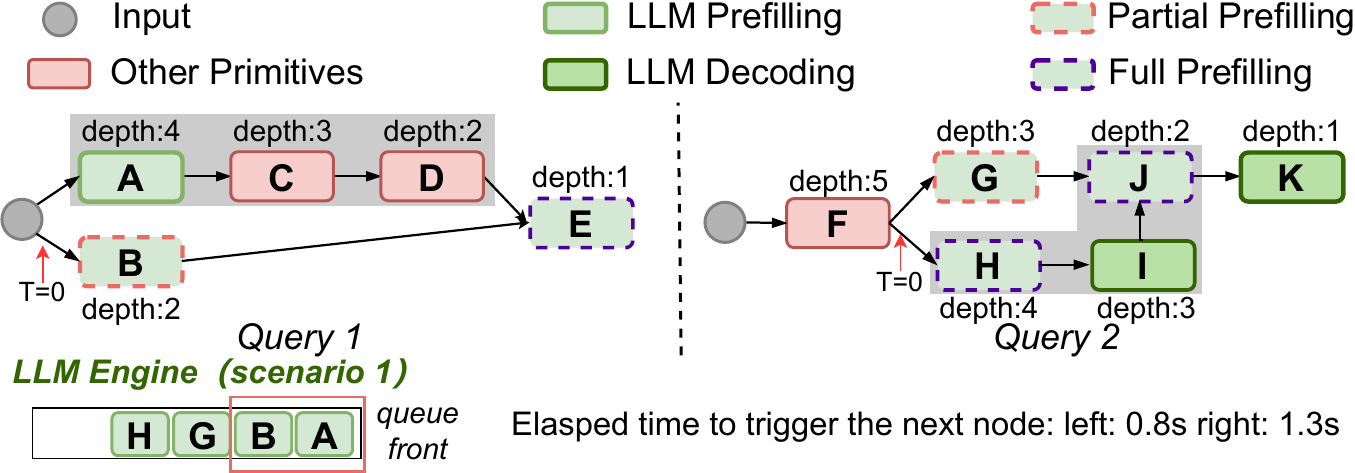}
        \caption{Blind batching.}
    \end{subfigure}
      \begin{subfigure}{\columnwidth}
        \centering
         \includegraphics[width=1.0\columnwidth]
    {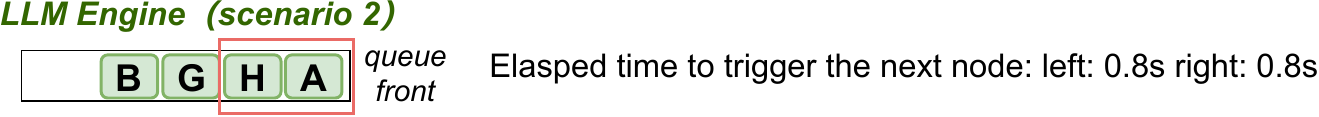}
        \caption{Topology-aware batching.}
    \end{subfigure}

     \caption{An illustrative comparison of two batching schemes for an LLM instance with a maximum token size of 1024. Assume each Prefilling or Partial/Full Prefilling input contains 512 tokens, with a latency of 0.5s for 512 tokens and 0.8s for a batch of two 512-token input.}
  \label{fig:batch} 
\end{figure}

\subsection{Engine Scheduler}\label{subsec:pscheduler}

Execution engine instances are managed by dedicated engine schedulers, enabling independent execution of primitive nodes mapped to different engine types. The main challenge is efficiently fusing primitives that request the same engine. With an optimized e-graph, a query may dispatch multiple primitive nodes simultaneously to an engine scheduler or have several pending primitive nodes in the queue, especially when components share the same engine, such as the proxy and judge modules in Figure~\ref{fig:app1} or the query expansion and LLM synthesizing modules in Figure~\ref{fig:app4} using the same LLM.

\noindent\textbf{Strawman solution and limitation: blind batching.} A naive approach to handling diverse primitive nodes is to treat them uniformly. A engine scheduler dynamically batches associated primitive requests from the pending queue using a FIFO policy, reaching a predefined maximum batch size or upon timeout, similar to existing systems~\cite{crankshaw2017clipper, tritonserver}. The batch is then dispatched to an engine instance. However, this simplistic method overlooks that not all primitive nodes from the same query equally contribute to graph progression.

As illustrated in Figure~\ref{fig:batch}, for query 1, primitive A and B requesting the LLM engine enter the queue along with primitive G and H from query 2. Blind batching would batch A and B, leaving G and H to wait. However, executing primitive B at this point yields little benefit since B's child E cannot be issued later due to E's other untriggered parent D. In contrast, batching A and H advances both queries' graph execution, with B's delay not bottlenecking query 1.

\begin{algorithm}[t]
\footnotesize
\caption{Topology-aware batching }\label{algo}
\begin{algorithmic}[1]
\State \textbf{Event 1: After getting the optimized e-graph $\mathcal{G}$ for a query:}
\Statex {\textcolor{blue}{\# \textit{Determine node's depth following a reversed topological sort.}}}
\State $\mathcal{G^\prime} \gets \texttt{RevTopoSort}(\mathcal{G})$; $\texttt{InitDepth}(\mathcal{G^\prime})$
\For{$v \in \mathcal{G^\prime}$}
\For{$p \in v.parents$}
 \State $p.depth \gets \max(p.depth, v.depth + 1)$
    \EndFor
\EndFor

\State \textbf{Event 2: On the scheduling period of an Engine Scheduler:}
\State {\textit{\textcolor{blue}{\# Form the batch based on primitives' depth and relationship}}}
\State $max\_bs \gets \texttt{GetConfigBatchsize()}$; $batch \gets []$
\State $\mathcal{B} \gets$ group nodes from the same query into buckets from the queue, sorted by the earliest arrival time of each bucket's node.
\For{$ {b} \in \mathcal{B}$}
    \State $slots \gets max\_bs-batch.size()$
    \If{$slots = 0$}
    \State break
    \EndIf
    \State $candidates \gets$ pop associated requests (up to $slots$) from each node with highest depth in $b$. 
    \State remove the nodes whose all associated requests are scheduled.
    \State $batch.append(candidates)$
\EndFor
\end{algorithmic}
\label{algo::topo_batch}
\end{algorithm}

\noindent\textbf{Our solution: topology-aware batching.} The example highlights the limitations of blind batching, which ignores the unique contributions of each primitive to the query's graph progression. Primitive nodes in a graph vary in topological depth; delaying lower-depth nodes can reserve resources for more contributive ones, enhancing overall execution. Besides, it is essential to consider the correlation and dependency between requests, unlike the approach taken by existing orchestration (as discussed in ~\Cref{subsec:moti_application_aware}). Combining these insights, we propose topology-aware batching, a heuristic solution that \textit{leverages the depth of primitive nodes and their relationships to intelligently guide batch formation}.

Concretely, the approach offers two primary benefits. First, for an individual query, depth information naturally captures the dependency among different primitives, enabling straightforward adjustments to the scheduling preferences with the inherent request correlation in each primitive (see \Cref{subsec:moti_application_aware}). For example, primitives at the same depth can be executed at the maximum efficient batch size to optimize throughput and advance the graph. Second, while depth information may not pinpoint the exact critical path due to unpredictable latency in real execution, it guides primitive prioritization for a query (see Figure~\ref{fig:batch}), facilitating efficient resource utilization across multiple queries.

Algorithm~\ref{algo::topo_batch} shows the procedure for topology-aware batching. After obtaining a query's e-graph, primitive nodes are reverse topologically sorted and their depths recorded, with the output node having the smallest depth (Event 1). When scheduling, primitives from the same query in the engine scheduler's queue are grouped into buckets. Within each bucket, primitives are sorted by depth, prioritizing those with higher depths. The buckets are then sorted based on the earliest arrival time within each bucket. Given a slot number, which represents the pre-determined maximum batch size (or maximum token size for LLM) that ensures optimal throughput efficiency, the scheduler iterates through each bucket. For each bucket, it examines the highest-priority primitive nodes and, if free slots are available, moves the associated requests from a primitive into the candidates (Event 2).

\subsection{Co-located Applications}\label{subsec:multi_app}

\sys is designed to optimize individual app workflows. 
When applications are co-located in the same infrastructure~\cite{zhang2023shepherd,gujarati2020clockwork}, they are orchestrated by \sys independently. 
That is, the e-graphs of different applications containing different or identical primitives with distinct app-level and query-level metadata, are treated by \sys in the same manner as in single-app scenarios. 
This leaves out the potential for further optimization when for example some co-located applications share the LLM engines and the model so advanced techniques like KV cache sharing across different applications may further improve throughput.
We plan to explore these opportunities in future work. 
For completeness, we evaluate \sys's capability to handle queries in co-located app scenarios with a discussion in~\Cref{subsec:eval-multi-app}.

\section{Implementation}
We implement the prototype of \sys with \textasciitilde5,300 lines of code in Python. Specifically, we leverage several existing libraries: (1) Ray~\cite{moritz2018ray} for distributed scheduling and execution; (2) LlamaIndex~\cite{llamaindex} for pre-processing tasks, such as text chunking and HTML/PDF parsing; (3) postgresql~\cite{postgresql} as the default database; (4) pgvector~\cite{pgvector} as the vector search engine; (5) Google custom search~\cite{googlesearch} as the search engine, supporting both single and batched requests; and (6) vLLM~\cite{kwon2023vllm} as the LLM serving engine, which we additionally modify to support \texttt{Partial Prefilling} and \texttt{Full Prefilling} in Table~\ref{tab:TPr_example}.

For the frontend, we provide user interfaces via FastAPI~\cite{Fastapi} for submitting queries and user configurations.
For the backend, the graph scheduler maintains a thread pool to allocate a dedicated thread for each new query, in order to construct, optimize and dispatch the e-graph. Beyond the discussion in ~\Cref{sec:runtime}, each engine scheduler also manages load balancing across different instances based on various load metrics -- primarily the number of executed requests for general engines and the occupied KV cache slots for LLMs.

\noindent\textbf{Mitigating communication overhead.} To reduce communication overhead in a central scheduler, we use a dependent pre-scheduling mechanism for adjacent primitives with large data interactions or the same execution engine. This allows simultaneous issuance of two dependent primitives, namely A and B, with B waiting for the output of A. Along with sending A's result to the scheduler, an RPC call also sends the output of A directly to the execution engine of B. This avoids relaying results through the scheduler before issuing B, and hence can reduce the communication overhead.

\section{Evaluation}
\begin{figure*}[t]
  \centering
  \includegraphics[width=0.98\textwidth]{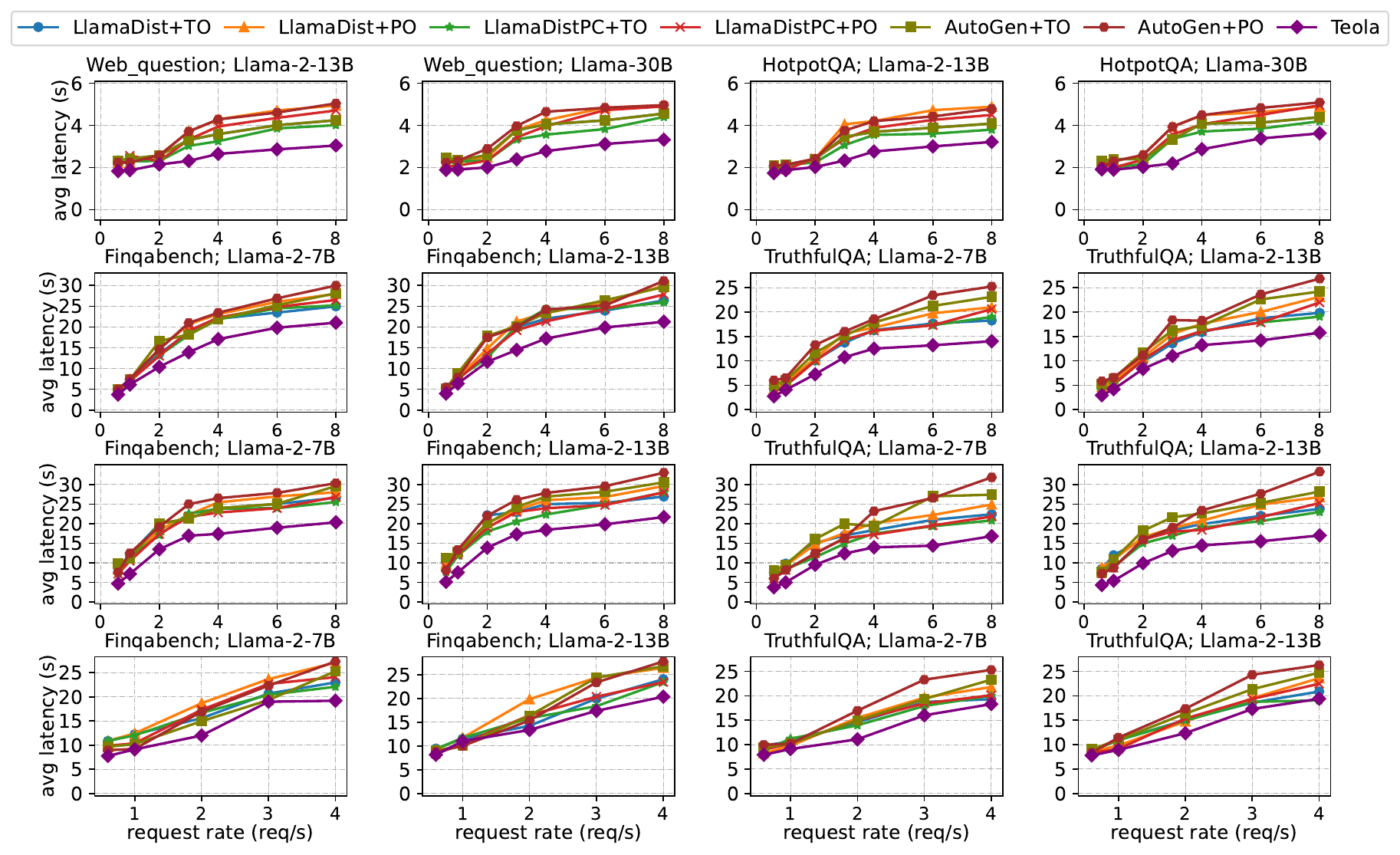} 
  \caption{{End-to-end performance of search engine-empowered generation (1st row), document QA with naive RAG (2nd row), document QA with advanced RAG (3rd row) and contextual retrieval (4th row). Subcaption of each subfigure indicates the dataset and core LLM.}}
  \label{fig:latency_performance} 
\end{figure*}

\noindent\textbf{Testbed setup.} We allocate model-based engines (e.g., LLMs) and model-free engines with GPUs and CPU-only resources, respectively. Each engine instance used for embeddings or other non-LLM models are each hosted on a single NVIDIA 3090 24GB GPU. {For LLMs, each instance of gemma-2-2B, llama-2-7B, and llama-2-13B~\cite{touvron2023llama2model} is deployed on 1, 1, and 2 NVIDIA 3090 GPUs, respectively.} Each instance of llama-30B~\cite{touvron2023llamamodel} is deployed on 2 NVIDIA A800 80GB GPUs. The network bandwidth between any physical servers is 100 Gbps.

\noindent\textbf{Baseline.} 
{
To our knowledge, few studies have specifically focused on optimizing LLM-based workflows execution in distributed settings. Therefore, we compare \sys with the following popular frameworks and their variants:
}

\begin{itemize}[leftmargin=*]
{\item \textit{LlamaDist}: a Ray-based distributed implementation of LlamaIndex (a popular framework with over 36k GitHub stars) that defines a chain of task modules to construct an application pipeline. Each task module invokes requests to different distributed backend engines. This implementation integrates Ray with LlamaIndex's orchestration approach, utilizing the same engines as \sys but differing in the granularity of orchestration.
}

\item \textit{LlamaDistPC (parallel \& cache-reuse)}: an advanced LlamaDist variant that examines the predefined pipeline and manually parallelizes independent modules for concurrent execution. It also incorporates prefix caching for LLM to avoid recomputation for partial instructions in the prompt, as proposed in some previous works~\cite{zheng2023sglang,liu2024optimizingquery,gim2023promptcache,lin2024parrot}.

{
\item \textit{AutoGen}: a popular framework for LLM applications with over 32k stars on GitHub. It constructs various agents for an application, with each agent managing several system modules (e.g., a retrieval agent handling indexing, query embedding, and searching) and communicating with each other following a pre-defined graph. 
}

\end{itemize}

For the request scheduling of deployed engines, we compare two approaches:
\begin{itemize}[leftmargin=*]
\item\textit{Per-Invocation oriented (PO)}: We slightly modify the invocation from the orchestration side and make requests in an invocation as a bundle (essentially adding extra correlation information that is exploited in \sys to enhance baselines). The engines schedule each bundle at a time, prioritizing latency preferences for each invocation.

\item\textit{Throughput oriented (TO)}: We pre-tune a maximum batch/token size for each engine (i.e., increasing the batch size for DNNs or token size for LLMs by powers of 2 until no further throughput gain is observed) and employ dynamic batching strategy~\cite{tritonserver,crankshaw2017clipper,kwon2023vllm}. This maximizes the overall throughput but ignores any relationships among requests.
\end{itemize}

\noindent\textbf{Applications, models and workloads.} Our experiments cover three applications:
\begin{itemize}[leftmargin=*]
   \item \textit{Search engine-empowered generation} (Figure~\ref{fig:app1}): A search engine assists a core LLM in generating answers. A smaller LLM (llama-2-7B) acts as a proxy and judge, formulating a heuristic answer and determining if a search is needed. Any search results (top 4 entities) are fed into the core LLM to synthesize the final answer. Workload requests are generated using a Poisson distribution-based synthesis of web\_question~\cite{webquestion} and HotpotQA~\cite{yang2018hotpotqa} datasets.

    \item \textit{Document QA with naive RAG} (Figure~\ref{fig:app3}): Users input documents or webpages along with the question. The app segments the documents into chunks (default size: 256, overlap:30), embeds them with the bge-large-en-v1.5 model~\cite{xiao2023bgeembedding}, and stores them in a vector database (postgresql \& pgvector). It retrieves the most relevant chunks (default: top 3) to generate a response with a \textit{tree}-based mode. The workload (i.e. question and documents/webpage) is synthesized from Finqabench~\cite{Finqabench} and TruthfulQA~\cite{lin2021truthfulqa} datasets using a Poisson distribution.

    \item \textit{Document QA with advanced RAG} (Figure~\ref{fig:app4}): Extending the second app, a query expansion is used (core LLM) to rewrite and expand the original query into multiple new queries (default: 3), improving retrieval accuracy. A reranker (bge-reranker-large~\cite{xiao2023bgeembedding}) evaluates the similarity between retrieved chunks, with each query searching for 16 chunks and determining the top 3 overall. These top chunks are fed into the core LLM for generation in a \textit{refine} mode as mentioned in \Cref{subsec:opt_example}.
{
    \item \textit{Contextual retrieval} (Figure~\ref{fig:app5}): Extending the naive RAG, this workflow incorporates contextualization using a lightweight LLM, gemma-2-2B~\cite{team2024gemma}, to process each chunk before indexing. To optimize time and cost, each chunk is contextualized with its four neighboring chunks, generating a concise context to prepend to each chunk. A reranker, identical to the one in the advanced RAG, reranks the 32 fetched chunks. The top 3 chunks are then input into the core LLM for one-shot generation.
}
    
\end{itemize}

Unless otherwise specified, the above default configurations are applied. All models are deployed in half precision, and different core LLMs (7B, 13B and 30B) are experimented.

\subsection{End-to-end Performance}
We evaluate the performance with different schemes for various apps, all under the same resource allocation. Each non-LLM engine is provisioned with a single instance, while each LLM is provisioned with two instances. 

{
\noindent\textbf{Search engine-empowered generation.} Figure~\ref{fig:latency_performance} (1st row) shows \sys outperforming the other four schemes by up to 1.79x. \sys's efficiency is attributed to parallelizable partial prefilling for instructions and questions for both the judge and core LLM, and effective batching coordination for different engines. In contrast, LlamaDist executes modules sequentially and struggles with request scheduling for multiple queries. PO's focus on per-invocation latency results in longer queue times under high request rates, while TO generally performs better in these scenarios. LlamaDistPC fails to benefit from parallelization due to the lack of explicit parallelization across modules. Its prefix caching for partial instructions (typically around 60 tokens) provides limited benefit, as prefix caching is most advantageous when prefixes are significantly longer~\cite{promptflow,zheng2023sglang}. AutoGen defines agents for the proxy model, judge model, search engine, and LLM synthesizer in this case, resulting in a structure similar to LlamaDist and therefore causing similar performance.

\noindent\textbf{Document QA with naive RAG.} Figure~\ref{fig:latency_performance} (2rd row) demonstrates that \sys outperforms the other four schemes by up to 1.62x at low request rates and 1.67x at high rates. LlamaDist executes modules sequentially without specific optimizations, while LlamaDistPC enables limited parallelization (indexing and query embedding modules) and partial instruction KV cache reuse, performing slightly better than LlamaDist. While AutoGen's workflow, consisting of a retrieval agent and synthesizer agent, performs adequately at low request volumes, its simplified orchestration architecture results in extended queuing delays when handling high request loads. Regarding scheduling, PO outperforms TO at low request rates due to its focus on per-invocation latency, but its performance suffers at high rates. Furthermore, the app introduces intricate request relationships, with both the indexing and query embedding modules utilizing the embedding model, and the LLM synthesizing module making three initial requests followed by a subsequent request to construct the \textit{tree} synthesis. If overlooked, these relationships can cause batching inefficiencies and reduced goodput, similar to TO. In contrast, \sys leverages the e-graph, incorporating pipelining to split compute-heavy tasks like large embeddings for document chunks, while also exploring more parallelization opportunities, such as four partial prefilling. Additionally, \sys's topology-aware batching captures dependencies and correlations among requests linked to different primitives, facilitating effective batching for each engine.

\noindent\textbf{Document QA with advanced RAG.} This app is the most complex in our settings, yet it provides ample opportunities to demonstrate the effectiveness of \sys. It leverages aggressive optimization techniques such as parallelization at different levels (e.g., independent dataflow branches and partial prefilling for different LLM calls) and pipelining (e.g., breaking large embeddings into smaller ones and splitting the decoding process in query expansion into three partial decodes), as shown in Figure~\ref{fig:opt_graph}. In contrast, LlamaDist runs sequentially with a simple run-to-completion paradigm, missing opportunities to reduce end-to-end latency. LlamaDistPC improves parallelization across the indexing and query expansion modules and reuses partial KV cache but still fails to explore the full optimization potential like \sys. For AutoGen, it would define four agents for retrieval, reranking, query expansion, and LLM synthesizer in this workflow, but it suffers from high request load due to its inability to pipeline and parallelize operations effectively. Additionally, similar to naive RAG, all other baselines struggle to efficiently coordinate requests whether in PO or TO, whereas \sys performs well. Overall, \sys outperforms others by up to 2.09x at low request rates and 2.03x at high request rates, as shown in Figure~\ref{fig:latency_performance} (3rd row).

\noindent\textbf{Contextual retrieval.} This workflow mainly adds contextualization with a lightweight LLM for chunks before indexing and a reranking step after searching, compared to naive RAG. However, due to the massive volume of chunks to process, the contextualization occupies a significant portion of time for all schemes. Although little graph-level optimization could be applied to the contextualization part, \sys's application-aware scheduling and graph optimization on other components still help it achieve 1.06x-1.59x speedup compared to other baselines, as shown in Figure~\ref{fig:latency_performance} (4th row), which further validates the applicability of \sys.
}

{

\subsection{Co-located Applications}\label{subsec:eval-multi-app}

\begin{figure}[t]
  \centering
  \includegraphics[width=0.48\textwidth]{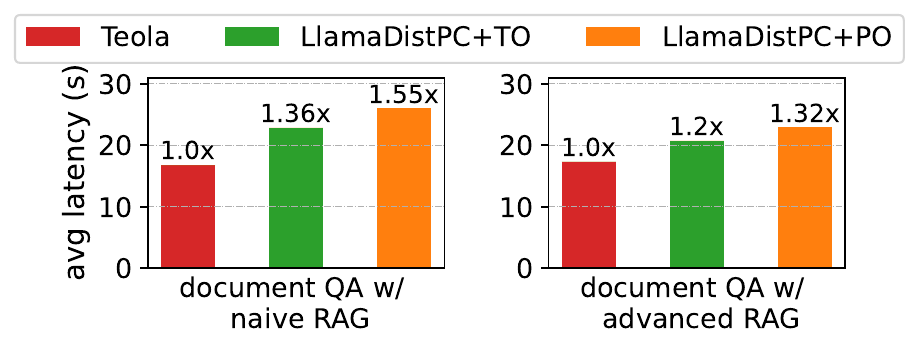} 
    \vspace{-4.5mm}
  \caption{{The average latency for two apps under a co-location scenario using llama-2-13B as the core LLM on truthfulQA dataset.}}
  \label{fig:eval_multiapp} 
\end{figure}

Beyond evaluating \sys's performance in single-application scenarios, we also assess its capability to handle concurrent queries across multiple applications. To demonstrate this, we present a scenario where two document QA applications—a naive RAG-based one and an advanced RAG-based one—operate simultaneously on the same underlying infrastructure. These apps share similar backend engines but differ in their calling semantics, with the advanced RAG-based QA requiring an additional reranking engine.

We compare the average latency of \sys against the stronger baseline, LlamaDistPC, at a request rate of 3 requests per second per application. Figure~\ref{fig:eval_multiapp} illustrates the average latency for queries from both applications, revealing that \sys maintains a performance advantage over LlamaDistPC in multi-app co-location scenarios. \sys achieves a 1.2x to 1.55x latency speedup across the two apps, highlighting the consistency of its capabilities in handling multiple application queries. This can be attributed to \sys's app-agnostic graph optimization and app-aware scheduling techniques.

Furthermore, in real-world deployments, applications could have varying priorities. For example, LLM chat services prioritize shorter latency to enhance user experience, while document summarization services may tolerate longer latencies. These differing priorities can influence the order of request processing within the infrastructure. The varying objectives among applications can be encoded as metadata within their generated primitives and used to facilitate priority-based scheduling or resource-constrained scheduling in each engine scheduler. However, a detailed exploration of this aspect is beyond the scope of this paper.

}

\subsection{Ablation Study}
We show the effectiveness of \sys's main components from graph optimization and runtime scheduling perspectives.

\begin{figure}[t]
  \centering
  \includegraphics[width=0.45\textwidth]{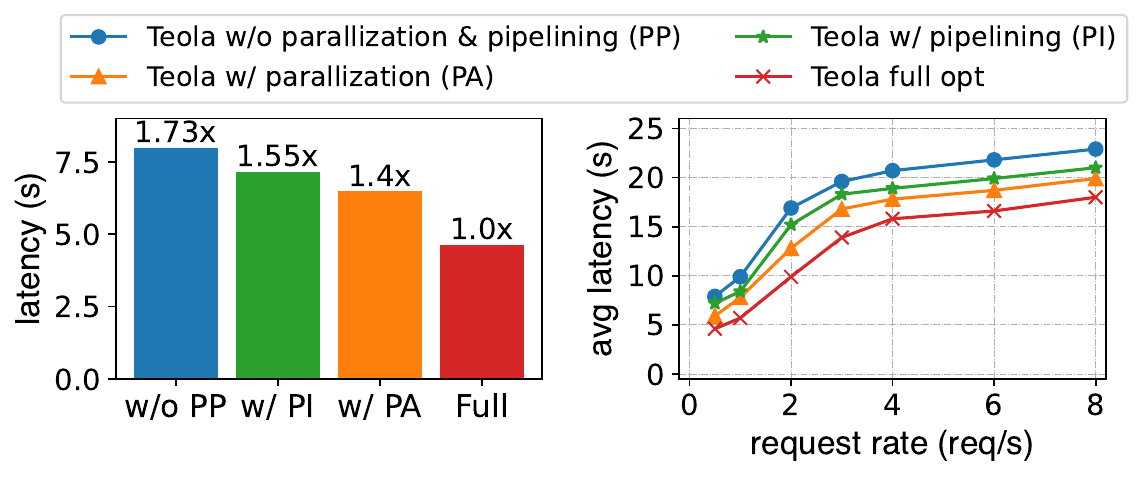} 
    \vspace{-3mm}
  \caption{Ablation study on graph optimization in document QA with advanced RAG on truthfulQA dataset using llama-30B as core LLM. Left: single-query latency averaged over 10 runs. Right: average latency under varying request loads.}
  \label{fig:ablation_graph_opt} 
\end{figure}

\noindent\textbf{Graph optimization.} As shown in Figure~\ref{fig:ablation_graph_opt}, we compare the performance of \sys under different scenarios, i.e., with or without parallelization (Pass 1 \& 3) and pipelining (Pass 2 \& 4) optimization as in~\Cref{sec:graph_opt}.
The left figure shows the average latency of a single query and explicitly demonstrates that both parallelization and pipelining effectively capture optimization opportunities and reduce latency. This holds true as well in a request trace with different request rates.

\begin{figure}[t]
    \centering
    \begin{minipage}[t]{0.280\textwidth}
       \centering
  \includegraphics[width=\textwidth]{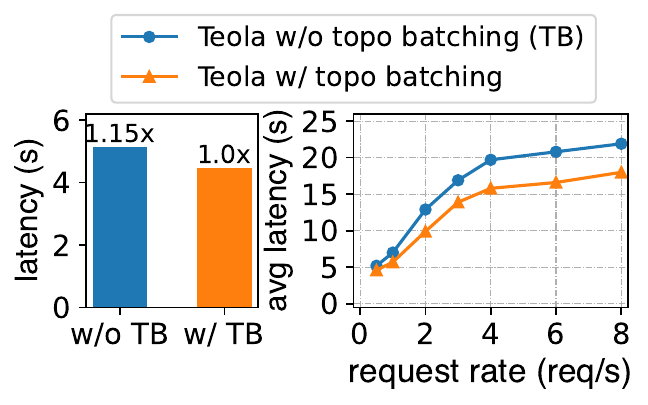} 
  \caption{Ablation study on runtime scheduling. The setting is same as that in Figure~\ref{fig:ablation_graph_opt}.}
  \label{fig:ablation_batching} 
    \end{minipage}
    \hfill
    \hspace{-0.25in}
    \begin{minipage}[t]{0.185\textwidth}
      \centering
  \includegraphics[width=\textwidth]{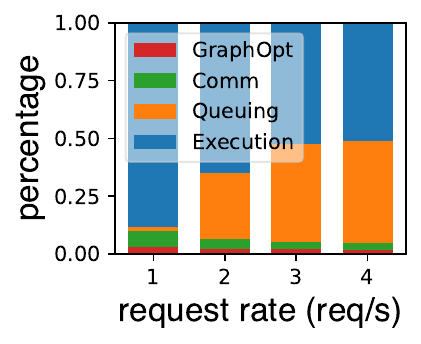} 
  \caption{Latency breakdown of \sys's execution critical path.}
  \label{fig:latency_breakdown} 
    \end{minipage}
   
\end{figure}

\noindent\textbf{Runtime scheduling.} Figure~\ref{fig:ablation_batching} illustrates the performance impact of enabling and disabling topology-aware batching as discussed in~\Cref{subsec:pscheduler}. The left figure demonstrates its effectiveness in capturing the varying depths of different primitives for the same engine and scheduling them with the informed correlation and dependency information to better meet the application-level performance. This results in an average 1.15x speedup for single query execution. In multi-query scenarios, topology-aware batching remains beneficial. Beyond single-query efficiency, it effectively fuses contributive primitives across queries, thereby facilitating overall execution and reducing average latency by up to 19.2\%.

\subsection{Overhead Analysis}

\begin{table}[t]
\resizebox{\columnwidth}{!}{%
\begin{tabular}{ccc|c}
\hline
Partial Prefilling  & Full Prefilling  & Total         & Single Prefilling  \\ \hline
76.03 (200)                & 215.89 (800)            & 291.92 (1000) & 260.36 (1000)             \\
217.67 (850)               & 222.66 (850)            & 440.33 (1700) & 414.09 (1700)             \\
582.95 (2500)              & 159.65 (500)            & 742.60 (3000) & 720.15 (3000)             \\ \hline
\end{tabular}%
}
\vspace{2mm}
 \caption{{Execution efficiency comparison between decomposed prefilling (partial and full prefilling) of \sys (left) and single complete prefilling (right) for different input sizes with llama-2-7B. Values represent execution time in millionsecond, with input size in parentheses.}}
\label{table:prefilling}
\end{table}

{

We have shown that \sys provides substantial performance improvements over existing solutions. 
Here in order to better understand \sys's overheads we provide a more detailed analysis in mainly three aspects: 
graph optimization, data communication, and execution efficiency.
First, e-graph optimization incurs new overhead for each query, but this is often minor due to the limited number of primitives in an LLM-based workflow and the ability to cache the results for queries of the same app. 
Second, communication across primitives also leads to higher compared to existing coarse-grained approach. 
\sys's pre-scheduling mechanism mitigates this, and the pipelined execution also allows partially conveyed data to be processed earlier. 
Third, decomposing certain operations, such as LLM prefilling, may compromise execution efficiency (e.g., repeatedly moving KV cache into SRAM or failing to fully utilize the GPU) compared to a single complete operation. Nevertheless, this performance sacrifice is minimal, as demonstrated by a comparison in Table~\ref{table:prefilling}, where the slowdown is only between 3.11\% and 12.12\%. Moreover, the gained parallelism further outweighs the slight loss and provides a larger end-to-end speedup.

We provide a latency breakdown for document QA with advanced RAG on the TruthfulQA dataset as a case study. 
We observe that graph optimization overhead is minimal, ranging from 1.3\% to 3\% of the total latency, when leveraging caching optimizations. Similarly, the communication overhead remains low, accounting for only 3.1\% to 6.2\%. Under increased request rates on fixed resources, queuing latency dominates end-to-end time, further diminishing the relative impact of these overheads on overall execution time.

}

\section{Limitations and Future Work.}\label{sec:discussion}
In this section, we point out some limitations of \sys and identify future research directions.

\noindent\textbf{Dynamic workflows.} 
LLM-based workflows typically require upfront human design but offer stable performance, leading to widespread adoption in both open-source projects~\cite{llamaindex,langchain,haystack,lazyllm} and production environments~\cite{alibaba_llmapp_observation,pairag,azurerag}. \sys is primarily designed for such structured scenarios. However, autonomous agents operating without human intervention have emerged, wherein the LLM plans and executes actions dynamically without a fixed workflow. 
\sys's ahead-of-time graph optimization would struggle with dynamic workflows, especially those involving self-reflection~\cite{madaan2024selfreflection}, as it cannot capture the complete primitive-level graph before execution. Adapting \sys for such dynamic workflows is a very promising avenue for future research.

\noindent\textbf{Coupling with the backends.} 
To enable finer-grained orchestration, we have to modify several engine-side mechanisms (e.g. support decomposed primitive operations and certain batching strategies), which requires extra engineering efforts compared to frameworks~\cite{llamaindex,langchain} that decouple the orchestration and execution and work with pluggable engines. 
This is an essential trade-off between performance and modularity. 
These optimization details remain encapsulated from the user's perspective.

\noindent\textbf{Exploitation of critical path.} Critical-path information in the e-graph can be further leveraged. For resource allocation, we can adjust resources for operations on critical and non-critical paths to maximize utilization based on workload patterns. For request scheduling, prioritizing critical nodes for specific queries can enhance the current topological batching, but this requires accurate online predictions of critical paths and coordination complexities.

\section{Related Work}\label{sec:related_work}

\noindent\textbf{LLM inference optimization.} LLM inference has garnered significant attention, with numerous studies focusing on various optimization directions, including kernel acceleration~\cite{xiao2023smoothquant,dao2022flashattention,hong2023flashdecoding++}, request scheduling~\cite{yu2022orca,agrawal2024taming,agrawal2023sarathi,sheng2023fairness}, model parallelism~\cite{li2023alpaserve,zhong2024distserve,miao2023spotserve}, semantic cache~\cite{bang2023gptcache,zhu2023optimalcache}, KV cache management~\cite{kwon2023vllm,wu2024loongserve,lin2024infinite}, KV cache reusing~\cite{kwon2023vllm,zheng2023sglang,gim2023promptcache,liu2024optimrelation,jin2024ragcache} and advanced decoding algorithms~\cite{ou2024losslessdecoding,liu2023onlinespec,miao2024specinfer}. Recent works~\cite{patel2023splitwise,zhong2024distserve,hu2024inferencewithoutinterference} disaggregate the deployment of the prefilling and decoding phases to increase goodput. This philosophy aligns well with \sys's decomposition approach and could be seamlessly integrated. While most works provide point solutions in the LLM domain, \sys takes a holistic view, optimizing the entire application workflow and facilitating cooperation among diverse components. Thus, the optimizations in LLM inference would complement \sys's efforts.

{
\noindent\textbf{Frameworks for LLM applications.} 
Apart from frameworks mentioned before \cite{llamaindex,langchain,promptflow}, several studies \cite{kim2023llmcompiler,khattab2023dspy,zheng2023sglang,lin2024parrot} focus on optimizing complex LLM tasks involving multiple LLM calls. 
They explore opportunities such as parallelism and KV cache sharing by developing specific programming interfaces or compilers. 
AI agent frameworks \cite{langgraph,shen2023hugginggpt,hong2023metagpt,hong2024data} enable LLMs to make decisions, select tools, and interact with other LLMs, reducing the need for human intervention but introducing specific challenges. 
\sys is more related to~\cite{llamaindex,langchain,promptflow}, working with an application of various components with a pre-defined workflow while focusing on end-to-end efficiency. 
Parrot \cite{lin2024parrot} also captures the application-level affinity of multiple LLM requests and facilitates joint scheduling. 
SGlang \cite{zheng2023sglang} explores the relationship between LLM requests and attempts to reuse the history KV cache efficiently. 
They share a similar spirit with us in exploiting opportunities beyond a single inference request.
The difference is also clear: \sys focuses on end-to-end optimization at the application level that involve both LLM and non-LLM components, which were not considered before.
}

{
\noindent\textbf{Data analytics systems.} 
Data analytics systems~\cite{isard2007dryad,zaharia2016apachespark,gog2015musketeer,bauer2012legion} employ various graph based dependency analysis to enhance distributed task execution of big data analytics workloads. 
\sys draws upon these techniques to perform its graph optimization.
However, these systems are designed for general-purpose data processing.
\sys introduces a new set of abstractions specifically designed for LLM-based workflows, which allows it to explore a much larger optimization space beyond graph optimization or those mentioned in this work.
}

\section{Conclusion}

We present \sys, a fine-grained orchestration framework for LLM-based applications. The core idea is orchestration using primitive-level dataflow graphs. This explicitly exposes the attributes of primitive operations and their interactions, enabling natural exploration of workflow-level optimizations for parallel execution. By leveraging the primitive relationships from the graph, \sys employs a topology-aware batching heuristic to intelligently fuse requests from primitives for execution. Testbed experiments demonstrate that \sys can outperform existing schemes across different applications.

\clearpage
\bibliographystyle{plain}

\bibliography{main}

\begin{thebibliography}{10}

\bibitem{llamaindex}
{LlamaIndex}.
\newblock \url{https://github.com/jerryjliu/llama_index}, 2022.

\bibitem{promptflow}
{Promptflow}.
\newblock \url{https://https://github.com/microsoft/promptflow}, 2023.

\bibitem{autogpt}
Autogpt.
\newblock \url{https://github.com/Significant-Gravitas/AutoGPT}, 2024.

\bibitem{bing}
{Bing Copilot}.
\newblock \url{https://www.bing.com/chat}, 2024.

\bibitem{characterai}
Character.ai/.
\newblock \url{https://character.ai/}, 2024.

\bibitem{contextual-retrieval}
contextual-retrieval.
\newblock \url{https://www.anthropic.com/news/contextual-retrieval}, 2024.

\bibitem{Fastapi}
Fastapi.
\newblock \url{https://fastapi.tiangolo.com/}, 2024.

\bibitem{Finqabench}
{Finqabench Dataset}.
\newblock \url{https://huggingface.co/datasets/lighthouzai/finqabench}, 2024.

\bibitem{googlesearch}
Google custom search.
\newblock \url{https://programmablesearchengine.google.com/}, 2024.

\bibitem{azurerag}
Gpt-rag.
\newblock \url{https://github.com/Azure/GPT-RAG}, 2024.

\bibitem{haystack}
haystack.
\newblock \url{https://github.com/deepset-ai/haystack}, 2024.

\bibitem{langchain}
Langchain.
\newblock \url{https://github.com/langchain-ai/langchain}, 2024.

\bibitem{langgraph}
{LangGraph}.
\newblock \url{https://python.langchain.com/docs/langgraph/}, 2024.

\bibitem{lazyllm}
Lazyllm.
\newblock \url{https://github.com/LazyAGI/LazyLLM}, 2024.

\bibitem{alibaba_llmapp_observation}
Observability of llm applications: Exploration and practice from the
  perspective of trace.
\newblock
  \url{https://www.alibabacloud.com/blog/observability-of-llm-applications-exploration-and-practice-from-the-perspective-of-trace_601604},
  2024.

\bibitem{openaifunc}
Openai function calling.
\newblock \url{https://platform.openai.com/docs/guides/function-calling}, 2024.

\bibitem{pairag}
Pairag.
\newblock \url{https://github.com/aigc-apps/PAI-RAG}, 2024.

\bibitem{perplexity}
Perplexity ai.
\newblock \url{https://www.perplexity.ai/}, 2024.

\bibitem{pgvector}
Pgvector.
\newblock \url{https://github.com/pgvector/pgvector}, 2024.

\bibitem{postgresql}
Postgresql.
\newblock \url{https://www.postgresql.org/}, 2024.

\bibitem{privatellm}
Privatellm.
\newblock \url{https://privatellm.app/en}, 2024.

\bibitem{tritonserver}
Triton inference server.
\newblock \url{https://github.com/triton-inference-server}, 2024.

\bibitem{tensorflow}
Mart{\'\i}n Abadi, Paul Barham, Jianmin Chen, Zhifeng Chen, Andy Davis, Jeffrey
  Dean, Matthieu Devin, Sanjay Ghemawat, Geoffrey Irving, Michael Isard,
  Manjunath Kudlur, Josh Levenberg, Rajat Monga, Sherry Moore, Derek~G. Murray,
  Benoit Steiner, Paul Tucker, Vijay Vasudevan, Pete Warden, Martin Wicke, Yuan
  Yu, and Xiaoqiang Zheng.
\newblock {TensorFlow}: A system for {Large-Scale} machine learning.
\newblock In {\em Proc.~USENIX OSDI}, 2016.

\bibitem{agrawal2024taming}
Amey Agrawal, Nitin Kedia, Ashish Panwar, Jayashree Mohan, Nipun Kwatra,
  Bhargav~S Gulavani, Alexey Tumanov, and Ramachandran Ramjee.
\newblock Taming throughput-latency tradeoff in llm inference with
  sarathi-serve.
\newblock {\em arXiv preprint arXiv:2403.02310}, 2024.

\bibitem{agrawal2023sarathi}
Amey Agrawal, Ashish Panwar, Jayashree Mohan, Nipun Kwatra, Bhargav~S Gulavani,
  and Ramachandran Ramjee.
\newblock Sarathi: Efficient llm inference by piggybacking decodes with chunked
  prefills.
\newblock {\em arXiv preprint arXiv:2308.16369}, 2023.

\bibitem{bang2023gptcache}
Fu~Bang.
\newblock Gptcache: An open-source semantic cache for llm applications enabling
  faster answers and cost savings.
\newblock In {\em Proc.~the 3rd Workshop for Natural Language Processing Open
  Source Software}, 2023.

\bibitem{bauer2012legion}
Michael Bauer, Sean Treichler, Elliott Slaughter, and Alex Aiken.
\newblock Legion: Expressing locality and independence with logical regions.
\newblock In {\em Proc. IEEE SC}, 2012.

\bibitem{webquestion}
Jonathan Berant, Andrew Chou, Roy Frostig, and Percy Liang.
\newblock Semantic parsing on {F}reebase from question-answer pairs.
\newblock In {\em Proc.~EMNLP}, October 2013.

\bibitem{crankshaw2017clipper}
Daniel Crankshaw, Xin Wang, Guilio Zhou, Michael~J Franklin, Joseph~E Gonzalez,
  and Ion Stoica.
\newblock Clipper: A $\{$Low-Latency$\}$ online prediction serving system.
\newblock In {\em Proc.~USENIX NSDI}, 2017.

\bibitem{dao2022flashattention}
Tri Dao, Dan Fu, Stefano Ermon, Atri Rudra, and Christopher R{\'e}.
\newblock Flashattention: Fast and memory-efficient exact attention with
  io-awareness.
\newblock In {\em Proc.~NeurIPS}, 2022.

\bibitem{devlin2018bert}
Jacob Devlin, Ming-Wei Chang, Kenton Lee, and Kristina Toutanova.
\newblock Bert: Pre-training of deep bidirectional transformers for language
  understanding.
\newblock In {\em Proc.~ACL}, 2018.

\bibitem{gao2023retrievalsurvey}
Yunfan Gao, Yun Xiong, Xinyu Gao, Kangxiang Jia, Jinliu Pan, Yuxi Bi, Yi~Dai,
  Jiawei Sun, and Haofen Wang.
\newblock Retrieval-augmented generation for large language models: A survey.
\newblock {\em arXiv preprint arXiv:2312.10997}, 2023.

\bibitem{gim2023promptcache}
In~Gim, Guojun Chen, Seung-seob Lee, Nikhil Sarda, Anurag Khandelwal, and Lin
  Zhong.
\newblock Prompt cache: Modular attention reuse for low-latency inference.
\newblock {\em arXiv preprint arXiv:2311.04934}, 2023.

\bibitem{gog2015musketeer}
Ionel Gog, Malte Schwarzkopf, Natacha Crooks, Matthew~P Grosvenor, Allen
  Clement, and Steven Hand.
\newblock Musketeer: all for one, one for all in data processing systems.
\newblock In {\em Proc. ACM Eurosys}, 2015.

\bibitem{gujarati2020clockwork}
Arpan Gujarati, Reza Karimi, Safya Alzayat, Wei Hao, Antoine Kaufmann, Ymir
  Vigfusson, and Jonathan Mace.
\newblock Serving $\{$DNNs$\}$ like clockwork: Performance predictability from
  the bottom up.
\newblock In {\em Proc.~USENIX OSDI}, 2020.

\bibitem{hong2023flashdecoding++}
Ke~Hong, Guohao Dai, Jiaming Xu, Qiuli Mao, Xiuhong Li, Jun Liu, Kangdi Chen,
  Hanyu Dong, and Yu~Wang.
\newblock Flashdecoding++: Faster large language model inference on gpus.
\newblock In {\em Proc.~Machine Learning and Systems}, 2023.

\bibitem{hong2024data}
Sirui Hong, Yizhang Lin, Bang Liu, Bangbang Liu, Binhao Wu, Danyang Li, Jiaqi
  Chen, Jiayi Zhang, Jinlin Wang, Li~Zhang, Lingyao Zhang, Min Yang, Mingchen
  Zhuge, Taicheng Guo, Tuo Zhou, Wei Tao, Wenyi Wang, Xiangru Tang, Xiangtao
  Lu, Xiawu Zheng, Xinbing Liang, Yaying Fei, Yuheng Cheng, Zongze Xu, and
  Chenglin Wu.
\newblock Data interpreter: An llm agent for data science.
\newblock {\em arXiv preprint arXiv:2402.18679}, 2024.

\bibitem{hong2023metagpt}
Sirui Hong, Mingchen Zhuge, Jonathan Chen, Xiawu Zheng, Yuheng Cheng, Ceyao
  Zhang, Jinlin Wang, Zili Wang, Steven Ka~Shing Yau, Zijuan Lin, Liyang Zhou,
  Chenyu Ran, Lingfeng Xiao, Chenglin Wu, and J{\"u}rgen Schmidhuber.
\newblock Metagpt: Meta programming for a multi-agent collaborative framework.
\newblock {\em arXiv preprint arXiv:2308.00352}, 2023.

\bibitem{hu2024inferencewithoutinterference}
Cunchen Hu, Heyang Huang, Liangliang Xu, Xusheng Chen, Jiang Xu, Shuang Chen,
  Hao Feng, Chenxi Wang, Sa~Wang, Yungang Bao, et~al.
\newblock Inference without interference: Disaggregate llm inference for mixed
  downstream workloads.
\newblock {\em arXiv preprint arXiv:2401.11181}, 2024.

\bibitem{huang2023Hallucination}
Lei Huang, Weijiang Yu, Weitao Ma, Weihong Zhong, Zhangyin Feng, Haotian Wang,
  Qianglong Chen, Weihua Peng, Xiaocheng Feng, Bing Qin, and Ting Liu.
\newblock A survey on hallucination in large language models: Principles,
  taxonomy, challenges, and open questions.
\newblock {\em arXiv preprint arXiv:2311.05232}, 2023.

\bibitem{huang2024tool}
Zhongzhen Huang, Kui Xue, Yongqi Fan, Linjie Mu, Ruoyu Liu, Tong Ruan, Shaoting
  Zhang, and Xiaofan Zhang.
\newblock Tool calling: Enhancing medication consultation via
  retrieval-augmented large language models.
\newblock {\em arXiv preprint arXiv:2404.17897}, 2024.

\bibitem{isard2007dryad}
Michael Isard, Mihai Budiu, Yuan Yu, Andrew Birrell, and Dennis Fetterly.
\newblock Dryad: distributed data-parallel programs from sequential building
  blocks.
\newblock In {\em Proc.~ACM Eurosys}, 2007.

\bibitem{jagerman2023queryexpansion}
Rolf Jagerman, Honglei Zhuang, Zhen Qin, Xuanhui Wang, and Michael Bendersky.
\newblock Query expansion by prompting large language models.
\newblock {\em arXiv preprint arXiv:2305.03653}, 2023.

\bibitem{jeong2024adaptiverag}
Soyeong Jeong, Jinheon Baek, Sukmin Cho, Sung~Ju Hwang, and Jong~C Park.
\newblock Adaptive-rag: Learning to adapt retrieval-augmented large language
  models through question complexity.
\newblock {\em arXiv preprint arXiv:2403.14403}, 2024.

\bibitem{jin2024ragcache}
Chao Jin, Zili Zhang, Xuanlin Jiang, Fangyue Liu, Xin Liu, Xuanzhe Liu, and Xin
  Jin.
\newblock Ragcache: Efficient knowledge caching for retrieval-augmented
  generation.
\newblock {\em arXiv preprint arXiv:2404.12457}, 2024.

\bibitem{khattab2023dspy}
Omar Khattab, Arnav Singhvi, Paridhi Maheshwari, Zhiyuan Zhang, Keshav
  Santhanam, Sri Vardhamanan, Saiful Haq, Ashutosh Sharma, Thomas~T Joshi,
  Hanna Moazam, et~al.
\newblock Dspy: Compiling declarative language model calls into self-improving
  pipelines.
\newblock {\em arXiv preprint arXiv:2310.03714}, 2023.

\bibitem{kim2023llmcompiler}
Sehoon Kim, Suhong Moon, Ryan Tabrizi, Nicholas Lee, Michael~W Mahoney, Kurt
  Keutzer, and Amir Gholami.
\newblock An llm compiler for parallel function calling.
\newblock {\em arXiv preprint arXiv:2312.04511}, 2023.

\bibitem{kwon2023vllm}
Woosuk Kwon, Zhuohan Li, Siyuan Zhuang, Ying Sheng, Lianmin Zheng, Cody~Hao Yu,
  Joseph Gonzalez, Hao Zhang, and Ion Stoica.
\newblock Efficient memory management for large language model serving with
  pagedattention.
\newblock In {\em Proc.~ACM SOSP}, 2023.

\bibitem{lewis2020retrieval}
Patrick Lewis, Ethan Perez, Aleksandra Piktus, Fabio Petroni, Vladimir
  Karpukhin, Naman Goyal, Heinrich K{\"u}ttler, Mike Lewis, Wen-tau Yih, Tim
  Rockt{\"a}schel, et~al.
\newblock Retrieval-augmented generation for knowledge-intensive nlp tasks,
  2020.

\bibitem{li2023alpaserve}
Zhuohan Li, Lianmin Zheng, Yinmin Zhong, Vincent Liu, Ying Sheng, Xin Jin,
  Yanping Huang, Zhifeng Chen, Hao Zhang, Joseph~E Gonzalez, et~al.
\newblock $\{$AlpaServe$\}$: Statistical multiplexing with model parallelism
  for deep learning serving.
\newblock In {\em Proc.~USENIX OSDI}, 2023.

\bibitem{lin2024infinite}
Bin Lin, Tao Peng, Chen Zhang, Minmin Sun, Lanbo Li, Hanyu Zhao, Wencong Xiao,
  Qi~Xu, Xiafei Qiu, Shen Li, et~al.
\newblock Infinite-llm: Efficient llm service for long context with
  distattention and distributed kvcache.
\newblock {\em arXiv preprint arXiv:2401.02669}, 2024.

\bibitem{lin2024parrot}
Chaofan Lin, Zhenhua Han, Chengruidong Zhang, Yuqing Yang, Fan Yang, Chen Chen,
  and Lili Qiu.
\newblock Parrot: Efficient serving of llm-based applications with semantic
  variable.
\newblock In {\em Proc.~USENIX OSDI}, 2024.

\bibitem{lin2021truthfulqa}
Stephanie Lin, Jacob Hilton, and Owain Evans.
\newblock Truthfulqa: Measuring how models mimic human falsehoods.
\newblock {\em arXiv preprint arXiv:2109.07958}, 2021.

\bibitem{liu2024optimizingquery}
Shu Liu, Asim Biswal, Audrey Cheng, Xiangxi Mo, Shiyi Cao, Joseph~E Gonzalez,
  Ion Stoica, and Matei Zaharia.
\newblock Optimizing llm queries in relational workloads.
\newblock {\em arXiv preprint arXiv:2403.05821}, 2024.

\bibitem{liu2024optimrelation}
Shu Liu, Asim Biswal, Audrey Cheng, Xiangxi Mo, Shiyi Cao, Joseph~E Gonzalez,
  Ion Stoica, and Matei Zaharia.
\newblock Optimizing llm queries in relational workloads.
\newblock {\em arXiv preprint arXiv:2403.05821}, 2024.

\bibitem{liu2023onlinespec}
Xiaoxuan Liu, Lanxiang Hu, Peter Bailis, Ion Stoica, Zhijie Deng, Alvin Cheung,
  and Hao Zhang.
\newblock Online speculative decoding.
\newblock {\em arXiv preprint arXiv:2310.07177}, 2023.

\bibitem{liu2024iterativere}
Yanming Liu, Xinyue Peng, Xuhong Zhang, Weihao Liu, Jianwei Yin, Jiannan Cao,
  and Tianyu Du.
\newblock Ra-isf: Learning to answer and understand from retrieval augmentation
  via iterative self-feedback.
\newblock {\em arXiv preprint arXiv:2403.06840}, 2024.

\bibitem{madaan2024selfreflection}
Aman Madaan, Niket Tandon, Prakhar Gupta, Skyler Hallinan, Luyu Gao, Sarah
  Wiegreffe, Uri Alon, Nouha Dziri, Shrimai Prabhumoye, Yiming Yang, et~al.
\newblock Self-refine: Iterative refinement with self-feedback.
\newblock In {\em Proc.~NeurIPS}, 2024.

\bibitem{miao2024specinfer}
Xupeng Miao, Gabriele Oliaro, Zhihao Zhang, Xinhao Cheng, Zeyu Wang, Zhengxin
  Zhang, Rae Ying~Yee Wong, Alan Zhu, Lijie Yang, Xiaoxiang Shi, et~al.
\newblock Specinfer: Accelerating large language model serving with tree-based
  speculative inference and verification.
\newblock In {\em Proc.~ACM ASPLOS}, 2024.

\bibitem{miao2023spotserve}
Xupeng Miao, Chunan Shi, Jiangfei Duan, Xiaoli Xi, Dahua Lin, Bin Cui, and
  Zhihao Jia.
\newblock Spotserve: Serving generative large language models on preemptible
  instances.
\newblock In {\em Proc.~ACM ASPLOS}, 2024.

\bibitem{moritz2018ray}
Philipp Moritz, Robert Nishihara, Stephanie Wang, Alexey Tumanov, Richard Liaw,
  Eric Liang, Melih Elibol, Zongheng Yang, William Paul, Michael~I Jordan,
  et~al.
\newblock Ray: A distributed framework for emerging $\{$AI$\}$ applications.
\newblock In {\em Proc.~USENIX OSDI}, 2018.

\bibitem{ou2024losslessdecoding}
Jie Ou, Yueming Chen, and Wenhong Tian.
\newblock Lossless acceleration of large language model via adaptive n-gram
  parallel decoding.
\newblock {\em arXiv preprint arXiv:2404.08698}, 2024.

\bibitem{patel2023splitwise}
Pratyush Patel, Esha Choukse, Chaojie Zhang, {\'I}{\~n}igo Goiri, Aashaka Shah,
  Saeed Maleki, and Ricardo Bianchini.
\newblock Splitwise: Efficient generative llm inference using phase splitting.
\newblock {\em arXiv preprint arXiv:2311.18677}, 2023.

\bibitem{shen2023hugginggpt}
Yongliang Shen, Kaitao Song, Xu~Tan, Dongsheng Li, Weiming Lu, and Yueting
  Zhuang.
\newblock Hugginggpt: Solving ai tasks with chatgpt and its friends in
  huggingface.
\newblock In {\em Proc.~NeurIPS}, 2023.

\bibitem{sheng2023fairness}
Ying Sheng, Shiyi Cao, Dacheng Li, Banghua Zhu, Zhuohan Li, Danyang Zhuo,
  Joseph~E Gonzalez, and Ion Stoica.
\newblock Fairness in serving large language models.
\newblock {\em arXiv preprint arXiv:2401.00588}, 2023.

\bibitem{tan2024small}
Jiejun Tan, Zhicheng Dou, Yutao Zhu, Peidong Guo, Kun Fang, and Ji-Rong Wen.
\newblock Small models, big insights: Leveraging slim proxy models to decide
  when and what to retrieve for llms.
\newblock {\em arXiv preprint arXiv:2402.12052}, 2024.

\bibitem{team2024gemma}
Gemma Team, Morgane Riviere, Shreya Pathak, Pier~Giuseppe Sessa, Cassidy
  Hardin, Surya Bhupatiraju, L{\'e}onard Hussenot, Thomas Mesnard, Bobak
  Shahriari, Alexandre Ram{\'e}, et~al.
\newblock Gemma 2: Improving open language models at a practical size.
\newblock {\em arXiv preprint arXiv:2408.00118}, 2024.

\bibitem{touvron2023llamamodel}
Hugo Touvron, Thibaut Lavril, Gautier Izacard, Xavier Martinet, Marie-Anne
  Lachaux, Timoth{\'e}e Lacroix, Baptiste Rozi{\`e}re, Naman Goyal, Eric
  Hambro, Faisal Azhar, et~al.
\newblock Llama: Open and efficient foundation language models.
\newblock {\em arXiv preprint arXiv:2302.13971}, 2023.

\bibitem{touvron2023llama2model}
Hugo Touvron, Louis Martin, Kevin Stone, Peter Albert, Amjad Almahairi, Yasmine
  Babaei, Nikolay Bashlykov, Soumya Batra, Prajjwal Bhargava, Shruti Bhosale,
  et~al.
\newblock Llama 2: Open foundation and fine-tuned chat models.
\newblock {\em arXiv preprint arXiv:2307.09288}, 2023.

\bibitem{attention}
Ashish Vaswani, Noam Shazeer, Niki Parmar, Jakob Uszkoreit, Llion Jones,
  Aidan~N Gomez, \L~ukasz Kaiser, and Illia Polosukhin.
\newblock {Attention is All you Need}.
\newblock In {\em Proc.~NeurIPS}, 2017.

\bibitem{wu2024loongserve}
Bingyang Wu, Shengyu Liu, Yinmin Zhong, Peng Sun, Xuanzhe Liu, and Xin Jin.
\newblock Loongserve: Efficiently serving long-context large language models
  with elastic sequence parallelism.
\newblock {\em arXiv preprint arXiv:2404.09526}, 2024.

\bibitem{wu2023autogen}
Qingyun Wu, Gagan Bansal, Jieyu Zhang, Yiran Wu, Shaokun Zhang, Erkang Zhu,
  Beibin Li, Li~Jiang, Xiaoyun Zhang, and Chi Wang.
\newblock Autogen: Enabling next-gen llm applications via multi-agent
  conversation framework.
\newblock {\em arXiv preprint arXiv:2308.08155}, 2023.

\bibitem{xiao2023smoothquant}
Guangxuan Xiao, Ji~Lin, Mickael Seznec, Hao Wu, Julien Demouth, and Song Han.
\newblock Smoothquant: Accurate and efficient post-training quantization for
  large language models.
\newblock In {\em Proc.~ICML}, 2023.

\bibitem{xiao2023bgeembedding}
Shitao Xiao, Zheng Liu, Peitian Zhang, and Niklas Muennighof.
\newblock C-pack: Packaged resources to advance general chinese embedding.
\newblock {\em arXiv preprint arXiv:2309.07597}, 2023.

\bibitem{yang2018hotpotqa}
Zhilin Yang, Peng Qi, Saizheng Zhang, Yoshua Bengio, William~W. Cohen, Ruslan
  Salakhutdinov, and Christopher~D. Manning.
\newblock {HotpotQA}: A dataset for diverse, explainable multi-hop question
  answering.
\newblock In {\em Proc.~EMNLP}, 2018.

\bibitem{yu2022orca}
Gyeong-In Yu, Joo~Seong Jeong, Geon-Woo Kim, Soojeong Kim, and Byung-Gon Chun.
\newblock Orca: A distributed serving system for $\{$Transformer-Based$\}$
  generative models.
\newblock In {\em Proc.~USENIX OSDI}, 2022.

\bibitem{zaharia2016apachespark}
Matei Zaharia, Reynold~S Xin, Patrick Wendell, Tathagata Das, Michael Armbrust,
  Ankur Dave, Xiangrui Meng, Josh Rosen, Shivaram Venkataraman, Michael~J
  Franklin, et~al.
\newblock Apache spark: a unified engine for big data processing.
\newblock {\em Communications of the ACM}, 2016.

\bibitem{zhang2023shepherd}
Hong Zhang, Yupeng Tang, Anurag Khandelwal, and Ion Stoica.
\newblock $\{$SHEPHERD$\}$: Serving $\{$DNNs$\}$ in the wild.
\newblock In {\em Proc.~USENIX NSDI}, 2023.

\bibitem{zheng2023sglang}
Lianmin Zheng, Liangsheng Yin, Zhiqiang Xie, Jeff Huang, Chuyue Sun, Cody~Hao
  Yu, Shiyi Cao, Christos Kozyrakis, Ion Stoica, Joseph~E Gonzalez, et~al.
\newblock Efficiently programming large language models using sglang.
\newblock {\em arXiv preprint arXiv:2312.07104}, 2023.

\bibitem{zhong2024distserve}
Yinmin Zhong, Shengyu Liu, Junda Chen, Jianbo Hu, Yibo Zhu, Xuanzhe Liu, Xin
  Jin, and Hao Zhang.
\newblock Distserve: Disaggregating prefill and decoding for goodput-optimized
  large language model serving.
\newblock {\em arXiv preprint arXiv:2401.09670}, 2024.

\bibitem{zhu2023optimalcache}
Banghua Zhu, Ying Sheng, Lianmin Zheng, Clark Barrett, Michael~I Jordan, and
  Jiantao Jiao.
\newblock On optimal caching and model multiplexing for large model inference.
\newblock {\em arXiv preprint arXiv:2306.02003}, 2023.

\end{thebibliography}

\end{document}